\documentstyle[12pt, epsf]{article}
\newcommand\ZZZ{{\hbox{ Z\kern-1.6mm Z}}}
\newcommand{\beq}{\begin{equation}}
\newcommand{\eeq}{\end{equation}}
\newcommand{\bea}{\begin{eqnarray}}
\newcommand{\eea}{\end{eqnarray}}
\newcommand{\ra}{\rangle}
\newcommand{\la}{\langle}
\newcommand{\lt}{\left}
\newcommand{\rt}{\right}
\newcommand{\Iop}{\relax{\rm I\kern-.18em I}}
\newcommand{\one}{{\hbox{ 1\kern-1.2mm l}}}

\newcommand{\g}{\gamma}
\newcommand{\gb}{\bar \gamma}
\newcommand{\Z}{{\bf Z}}
\newcommand{\mv}{M^{V}}

\newcommand{\al}{\alpha}
\newcommand{\bt}{\beta}
\newcommand{\dt}{\delta}

\newcommand{\del}{\partial}
\newcommand{\F}{{\cal F}}

\newcommand{\B}{{\cal B}}
\newcommand{\CH}{{\cal H}}

\newcommand{\zb}{\bar z}
\newcommand{\wb}{\bar w}

\newcommand{\eps}{\epsilon}
\newcommand{\M}{{\cal M}}

\newcommand{\lam}{\lambda}

\newcommand{\K}{{\cal K}}

\newcommand{\th}{\theta}

\newcommand{\Ut}{\tilde U}

\newcommand{\Vt}{\tilde V}

\newcommand{\sectiono}[1]{\section{#1}\setcounter{equation}{0}}
\newcommand{\subsectiono}[1]{\subsection{#1}}

\begin{document}
{}~
{}~
\hfill\vbox{\hbox{DAMTP-2005-128} \hbox{UK/05-15} \hbox{hep-th/0512161}}\break

\vskip .6cm

\centerline{\Large \bf DDF Construction and D-Brane Boundary States}
\medskip
\centerline{\Large \bf in Pure Spinor Formalism}

\medskip

\vspace*{4.0ex}

\centerline{\large \rm Partha Mukhopadhyay }

\vspace*{4.0ex}

\centerline{\large \it Department of Applied Mathematics and
Theoretical Physics}
\centerline{\large \it University of Cambridge}
\centerline{\large \it Wilberforce Road, Cambridge CB3 0WA, UK}

\vspace*{2ex}
\centerline{\large and }
\vspace*{2ex}

\centerline{\large \it  Department of Physics and Astronomy}
\centerline{\large \it  University of Kentucky}
\centerline{\large \it  Lexington, KY-40506, USA}

\medskip

\centerline{E-mail: P.Mukhopadhyay@damtp.cam.ac.uk}

\vspace*{5.0ex}

\centerline{\bf Abstract} \bigskip

Open string boundary conditions for non-BPS D-branes in type II string
theories discussed in hep-th/0505157 give rise to two sectors with
integer (R sector) and half-integer (NS sector) modes for the combined
fermionic matter and bosonic ghost variables in pure spinor
formalism. Exploiting the manifest supersymmetry of the formalism we
explicitly construct the DDF (Del Giudice, Di Vecchia, Fubini) states
in both the sectors which are in one-to-one correspondence with the
states in light-cone Green-Schwarz formalism. We also give a proof of
validity of this construction. A similar construction in the closed string
sector enables us to define a physical Hilbert space in pure spinor
formalism which is used to project the covariant boundary states of
both the BPS and non-BPS instantonic D-branes. These projected
boundary states take exactly the same form as those found in
light-cone Green-Schwarz formalism and are suitable for computing the cylinder
diagram with manifest open-closed duality.

\vfill \eject

\tableofcontents

\baselineskip=18pt

\sectiono{Introduction and Summary}
\label{s:intro}

It has been a long standing problem of fundamental interest to
quantise superstring theory with all the space-time symmetries
manifest until Berkovits' proposal of pure spinor formalism
\cite{purespinor} was put forward. The formalism comes with a bag of
tools which includes a conformally gauge fixed
action with total central charge
zero, a BRST operator, physical state condition and rules for
computing scattering amplitudes. It is remarkable that everything fits
together to give consistent results like spectrum of physical states
and super-Poincar\'e covariant results for the scattering amplitudes in
flat space \cite{cohomology, consistency, loop}\footnote{See also
\cite{other}.}.
Although the space-time symmetry is very much emphasised,
it is not very clear what role the world-sheet conformal symmetry has
to play in this case, something which is very transparent in the
Neveu-Schwarz-Ramond (NSR) formalism. A particular example is to
understand D-branes\footnote{D-branes have been studied from various points of 
view in pure spinor formalism in \cite{Dbranes, schiappa}.} 
 from both the open and closed string point of
view. In NSR formalism these two views are bridged by the underlying
world-sheet picture
where the modular transformation relates them. This picture is not, in
general, apparent if one is armed only with a BRST setup like in pure
spinor formalism. Let us consider the simplest diagram, namely the
cylinder, which computes the force between two D-branes. Following
will be the generic procedure to compute this in a BRST setup: start
out with the quadratic space-time action involving the BRST operator
with a linearised gauge invariance. Obtain the propagator by inverting
the kinetic term with a valid gauge choice. Then compute the relevant
Feynman diagram where two external sources are
connected by a single line (see, for example, \cite{polchinski}).
If we know the correct strength for all
the sources then this computation is completely well-defined, the only
problem being there are infinite number of fields to be taken into account.
In this computation we do not use any CFT techniques as there
is no world-sheet interpretation and therefore the open-closed duality
is not manifest. In NSR formalism this interpretation results from a
simple gauge choice \cite{witten86} which we call Siegel
gauge\footnote{Its bsonic counterpart is called Siegel gauge in string
  field theory analysis.}. It is in this gauge
the closed string propagator in Schwinger parametrisation has the
interpretation of world-sheet time evolution. It is not clear what
would be the corresponding gauge choice in pure spinor formalism.

In more technical terms the problem can be described in the following way.
The cylinder diagram is computed in the closed string channel by first
constructing the boundary state in Siegel gauge and then computing an
inner product where the world-sheet time evolution operator is
sandwiched between two boundary states. The result can then be
interpreted in the open string channel by performing a modular
transformation. There are two basic ingredients in this computation:
\begin{enumerate}
\item
A suitable boundary state that provides the correct source terms
for all the relevant closed string states.
\item
The correct choice of degrees of freedom that should be allowed to
propagate along the cylinder (which is implemented by the Siegel gauge).
\end{enumerate}
In pure spinor formalism the first problem can be solved without much
trouble. Although there is a pure spinor constraint on the bosonic
ghost sector which makes the construction of boundary states
troublesome \cite{schiappa}, it has been suggested in
\cite{mukhopadhyay05} that constructing the boundary states in the relaxed
CFT where there is no constraint also does the job. Writing down the
boundary conditions and boundary states in the free CFT simply
bypasses the technical difficulty of incorporating the pure spinor
constraint yet producing the correct results for the source terms once
the rules for such computations are set up properly \cite{mukhopadhyay05}.
The main point of doing this is the fact that it is not beneficial to go
through the technical difficulty of imposing the pure spinor
constraint as this, by itself, does not solve the second problem. To
solve the second problem one also needs to throw away the gauge
degrees of freedom. This claim has been explicitly demonstrated in
\cite{mukhopadhyay05} through a computation of the long range force
between two D-branes. If it requires us to gauge fix the
space-time theory at every mass level separately then that would be an
uncontrollable job to do. A full string theoretic treatment of allowing
only the correct degrees of freedom to propagate seems to be a subtle issue.

Certainly the above subtlety is encountered when one tries to do the
computation with full covariance under $SO(9,1)$. Here we shall show
that the cylinder diagram can indeed be computed in pure spinor
formalism more easily by preserving covariance only under the transverse $SO(8)$ part.
The approach will be as follows: since the computation is well understood
in the light-cone Green-Schwarz (LCGS) formalism, it will suffice us
to construct the LCGS boundary states \cite{green96, mukhopadhyay00,
nemani, mukhopadhyay04} explicitly in pure spinor formalism. In other
words if the LCGS Hilbert space can be constructed explicitly in pure
spinor formalism, then any covariant pure spinor boundary state could
be projected onto that Hilbert space. The projected boundary states
could then be evolved by the world-sheet time evolution. By exploiting
the manifest space-time supersymmetry of pure spinor formalism we
construct the LCGS Hilbert space, which will be denoted $\CH_{DDF}$,
by going through the analogue of well-known DDF (Del Giudice,
Di Vecchia, Fubini) construction \cite{DDF}.

For open strings on a BPS D-brane this construction is done by first
using ghost number one, dimension zero unintegrated massless vertex
operators to construct certain massless physical states in the vector
and conjugate spinor representations of $SO(8)$ with special
kinematical condition that the light-cone component $q^+$ of the
momentum is non-zero and fixed. Then we use the ghost number zero,
dimension one integrated massless vertex operators to construct the
DDF operators which are the analogues of the LCGS oscillators. These
operators commute/anticommute with the BRST operator so that while acting on
physical states they produce other physical states. The DDF operators
constructed this way have nontrivial expansions in terms of the
fermionic matter variable $\th$. Therefore, although the leading
contribution to the DDF commutation relations do match with that of the
LCGS oscillators \cite{GSW}, there are terms higher order in $\th$. We
define the physical Hilbert space ${\cal H}_{DDF}$ to be
spanned by all the states which are obtained by applying creation
modes on the massless physical states constructed in the first
step. The ghost number two conjugate states are similarly constructed
by applying DDF operators on certain massless states that form the BRST
cohomology at ghost number 2. These states are chosen so that they are
conjugate to the ghost number one massless DDF states. We prove that
the DDF states constructed this way form an orthonormal basis in
${\cal H}_{DDF}$. The orthogonality of the DDF states establishes the
fact that all the higher order $\th$-terms drop off when the DDF
commutators are restricted in ${\cal H}_{DDF}$, so that the commutation
relations exactly match with those of LCGS oscillators. This implies
that though the DDF operators constructed here have complicated
$\th$-expansions they behave as simply as LCGS oscillators in $\CH_{DDF}$.

For a non-BPS D-brane, we have argued, using the boundary conditions
suggested in \cite{mukhopadhyay05}, that there are two sectors of open
strings - R and NS sectors. All the world-sheet fields that are
space-time fermions (i.e. fermionic matter and bosonic ghost) satisfy
periodic and anti-periodic boundary conditions on the doubled surface
in R and NS sectors respectively. DDF construction for the R sector
goes through as described in the previous paragraph. For the NS sector
the bosonic DDF operators are constructed in the same way. But the
fermionic ones are constructed in a slightly different way so that
they have half-integer modes instead of integer modes. This sector has
a unique ground state which is included in the BRST cohomology. We
identify this with the open string tachyon. 
This way the DDF states in the NS sector gives an
explicit construction of the corresponding open string spectrum found
in LCGS formalism \cite{yoneya99, mukhopadhyay04}.

Doing the similar construction on the closed string sector we define
the physical Hilbert space ${\cal H}_{DDF}$ on the closed string
side. Given the covariant boundary states constructed in the free CFT, as in
\cite{mukhopadhyay05}, projection of the corresponding actual pure
spinor boundary states onto ${\cal H}_{DDF}$ can be constructed
unambiguously. Due to the special kinematical condition of the DDF
states that all of them have a fixed nonzero $q^+$, only instantonic
D-brane boundary states can be projected this way to get the physical
components. To practically derive a projected boundary state we
proceed as follows. Since a projected boundary state is supposed to be expanded
in terms of the DDF states, it should be possible to get this as a
solution to the gluing conditions satisfied by the DDF
operators. Starting from the boundary conditions written in the open string
channel we derive the DDF gluing conditions and show that they are
given by the same equations satisfied by the LCGS
oscillators as discussed in \cite{green96, nemani, mukhopadhyay04}.
Therefore the projected boundary states in pure spinor formalism are
obtained from the corresponding boundary states in LCGS formalism
simply by interpreting the LCGS oscillators as the DDF operators
constructed here. These boundary states can then be evolved by the
world-sheet time evolution in pure spinor formalism to give the
correct result for the cylinder. In all our discussion we shall
consider type IIB string theory for definiteness and work in the
$\alpha^{\prime}=2$ unit. Generalisation to type IIA is straightforward.

The rest of the paper is organised as follows: section \ref{s:nbps}
reviews the non-BPS boundary conditions as suggested in
\cite{mukhopadhyay05} and analyses the open string spectrum. Section
\ref{s:DDF} discusses the DDF construction for open strings on both
BPS and non-BPS D-branes of type II string theories. This also includes a
proof of validity of the construction. Section \ref{s:physical}
defines the projected boundary states and derives the DDF gluing
conditions. The line of argument to compute the cylinder diagram has
been given in section \ref{s:cylinder}. We conclude with a few
unresolved questions in section \ref{s:conclusion}. Several appendices
contain necessary technical details.

\sectiono{Boundary Conditions and Spectrum of Open Strings on non-BPS D-branes}
\label{s:nbps}

\subsectiono{Review of Boundary Conditions}
\label{ss:bc}

$SO(8)$ covariant open string boundary conditions for non-BPS D-branes in
LCGS formalism were obtained in \cite{mukhopadhyay04}. Generalising
this work to any manifestly supersymmetric formalism, similar boundary
conditions were suggested in pure spinor formalism in
\cite{mukhopadhyay05}. Specialising to type IIB string theory, these
boundary conditions take the following form for the combined fermionic
matter and bosonic ghost sector in the unconstrained CFT,
\bea
\lt. \begin{array}{l}
U^{\al}(z)U^{\bt T}(w) = \M^{\al \bt}_{\g \dt}
\Ut^{\g}(\zb) \Ut^{\dt T}(\wb)~, \\ \\
U^{\al}(z)V_{\bt}^T(w) = \M^{\al~~\dt}_{~\bt \g}
\Ut^{\g}(\zb) \Vt_{\dt}^T(\wb)~, \\ \\
V_{\al}(z)V_{\bt}^T(w) = \M^{\g \dt}_{\al \bt}
\Vt_{\g}(\zb) \Vt_{\dt}^T(\wb)~,
\end{array} \rt\} \hbox{at } z=\zb, w=\wb ~,
\label{bcnBPS}
\eea
where we have introduced the column vectors,
\bea
U^{\al}(z) = \pmatrix{\lam^{\al}(z) \cr \th^{\al}(z)}~, ~~
V_{\al}(z) = \pmatrix{w_{\al}(z) \cr p_{\al}(z)}~,
\label{UV}
\eea
and similarly for the right moving sector. The coupling matrices are given by,
\bea
\M^{\al \bt}_{\g \dt} &=& -\lt[ {1\over 16} \g_{\mu}^{\al \bt}
(\mv)^{\mu}_{~\nu} \gb^{\nu}_{\g \dt} + {1\over 16\times 3!}
\g_{\mu_1 \cdots \mu_3}^{\al \bt} (\mv)^{\mu_1}_{~\nu_1}\cdots
(\mv)^{\mu_3}_{~\nu_3} \gb^{\nu_1\cdots \nu_3}_{\g \dt} \rt. \cr &&
\lt. +
{1\over 16\times 5!} \sum_{\mu_1,\cdots ,\mu_5 \in \K^{(5)}}
\g_{\mu_1\cdots \mu_5}^{\al \bt} (\mv)^{\mu_1}_{~\nu_1} \cdots
(\mv)^{\mu_5}_{~\nu_5} \gb^{\nu_1\cdots \nu_5}_{\g \dt} \rt]~, \cr
\M^{\al~~\dt}_{~\bt \g} &=& {1\over 16} \dt^{\al}_{~\bt}
\dt_{\g}^{~\dt} + {1\over 16 \times 2!}
\g_{\mu_1\mu_2~\bt}^{~~~~\al} (\mv)^{\mu_1}_{~\nu_1}
(\mv)^{\mu_2}_{~\nu_2} \gb^{\nu_1\nu_2~\dt}_{~~~~\g} \cr && +{1\over
16 \times 4!} \g_{\mu_1\cdots \mu_4~\bt}^{~~~~~~\al}
(\mv)^{\mu_1}_{~\nu_1} \cdots (\mv)^{\mu_4}_{~\nu_4}
\gb^{\nu_1\cdots \nu_4~\dt}_{~~~~~~\g}~.
\label{calMs}
\eea
Our gamma matrix conventions can be found in appendix
\ref{a:notation}. The summation convention for the repeated indices
has been followed
for all the terms in the above two equations except for the last term
of the first equation. The sum over the five vector indices $\mu_1
\cdots \mu_5$ has been restricted to a set $\K^{(5)}$ which is
defined as follows. We divide the set of all possible sets of five
indices $\{\{\mu_1,\cdots,\mu_5\}|\mu_i =0,\cdots ,9\}$ into two
subsets of equal order, namely $\K^{(5)}$ and $\K^{(5)}_D$ such that for every
element $\{\mu_1,\cdots,\mu_5\} \in \K^{(5)}$ there exists a dual element
$\{\mu_1,\cdots,\mu_5\}_D=\{\nu_1,\cdots,\nu_5\} \in \K^{(5)}_D$ such that,
$\eps^{\mu_1\cdots \mu_5\nu_1\cdots \nu_5} \neq 0$. The supersymmetry
currents, being odd in the world-sheet fields belonging to the
combined fermionic matter and bosonic ghosts, do not satisfy a linear
boundary condition. But the above boundary conditions do lead to BRST
invariance.

Following the method of \cite{mukhopadhyay04} we can now introduce the
holomorphic fields ${\cal U}^{\al}(z)$ and ${\cal V}_{\beta}(z)$ on
the doubled surface through the following expressions,
\bea
{\cal U}^{\al}(u)\cdots {\cal U}^{\bt T}(v) &=& \lt \{
\begin{array}{ll}
{\cal U}^{\al}(z)\cdots {\cal U}^{\bt T}(w)|_{z=u, w=v}~, &
\Im u, \Im v \geq 0~, \\
\M^{\al \bt}_{\g \dt}
\tilde {\cal U}^{\g}(\zb) \cdots \tilde {\cal U}^{\dt
  T}(\wb)|_{\zb=u,\wb=v}~, & \Im u, \Im v \leq 0~,
\end{array}  \rt. \\ && \cr
{\cal U}^{\al}(u)\cdots {\cal V}_{\bt}^T(v) &=& \lt \{
\begin{array}{ll}
{\cal U}^{\al}(z)\cdots {\cal V}_{\bt}^T(w)|_{z=u, w=v}~, &
\Im u, \Im v \geq 0~, \\
\M^{\al~~\dt}_{~\bt \g}
\tilde {\cal U}^{\g}(\zb) \cdots \tilde {\cal
  V}_{\dt}^T(\wb)|_{\zb=u,\wb=v}~, & \Im u, \Im v \leq 0~,
\end{array} \rt. \\ && \cr
{\cal V}_{\al}(u)\cdots {\cal V}_{\bt}^T(v) &=& \lt \{
\begin{array}{ll}
{\cal V}_{\al}(z) \cdots {\cal V}_{\bt}^T(w)|_{z=u, w=v}~, &
\Im u, \Im v \geq 0~, \\
\M^{\g \dt}_{\al \bt}
\tilde {\cal V}_{\g}(\zb) \cdots \tilde {\cal
V}_{\dt}^T(\wb)|_{\zb=u,\wb=v}~, ~~& \Im u, \Im v \leq 0~.
\end{array} \rt.
\label{calUV}
\eea
The dots imply that the relations are considered to be true even when
other operators appear in between in a correlation function. As argued
in \cite{mukhopadhyay04}, an immediate consequence of the above
definitions is,
\bea
{\cal U}^{\al}(\tau, 2\pi) = \pm {\cal U}^{\al}(\tau, 0)~, \quad
{\cal V}_{\al}(\tau ,2\pi) = \pm {\cal V}_{\al}(\tau, 0)~,
\label{periodicity}
\eea
so that there are two sectors of open strings, namely the R (periodic)
and the NS (anti-periodic) sectors. Below we shall discuss the
spectrum of these open strings.

\subsectiono{Open String Spectrum}
\label{ss:spectrum}

The periodic sector can be analysed in the usual way \cite{purespinor}
and therefore will give rise to an open string spectrum which is same
as that on a BPS D-brane of same dimensionality in type IIA
theory. Therefore the anti-periodic sector will be our main topic of
discussion here. In this sector all the relevant fields have half-integer
modes: ${\cal U}^{\al}_r$ and ${\cal V}_{\al,r}$, with $r\in \Z
+1/2$. Therefore, in absence of any zero modes, there is a unique
ground state $|\sigma \ra$ in the combined fermionic matter and
bosonic ghost sector, defined in the following way,
\bea
{\cal U}^{\al}_r |\sigma \ra = 0 ~, \quad
{\cal V}_{\al,r} |\sigma \ra = 0~, \quad \quad \forall r \geq 1/2~.
\label{sigma-def}
\eea
We may define the ghost number for this state to be one.
Excited states in the theory are obtained by applying the negative
modes of the oscillators on $|\sigma \ra$ and by imposing pure spinor
constraint. Physical states are the ghost number one states in BRST
cohomology, where the BRST operator (see eq.(\ref{QB})) takes the following
form in terms of various modes in $\alpha^{\prime}=2$ unit,
\bea
Q_B= \sum_r \lt(\lambda_{-r} p_r\rt) + {i\over 2} \sum_{r,s}
\lt(\lambda_{-r}\gb_{\mu}\theta_{-s}\rt) \alpha^{\mu}_{r+s}
-{1\over 8} \sum_{r,s,t} (r+s+t) \lt(\lambda_r \gb^{\mu}
\theta_s\rt) \lt(\theta_t \gb_{\mu}\theta_{-r-s-t}\rt)~,
\label{QB-modes}
\eea
where $\alpha^{\mu}_n=\oint {dz\over 2\pi} z^n \partial X^{\mu}(z)$.
Mass of the state is determined by the fact that the state is
annihilated by the Virasoro zero mode $L_0$ given in
eq.(\ref{L0}). It is easy to check that the ground state $|\sigma,
k\ra$, 
with momentum $k$, is an allowed state in the BRST cohomology.
\bea
Q_B |\sigma, k\ra = 0~, ~(\hbox{for any } k)~, \quad \quad
L_0 |\sigma, k\ra =0~, \Rightarrow   M^2 = -{1\over 4}~.
\eea
This is the open string tachyon.
There is a space-time fermion at the massless level. The
chirality is same as that of the massless fermion in the
R-sector\footnote{The
correct way to determine the chirality is to first realise that
these massless fermions are the goldstinos corresponding to the
spontaneously broken space-time supersymmetries \cite{yoneya99}. This
tells us that they should have opposite chirality in type IIA theory
which, in turn, determines the chirality of fermionic matter and
bosonic ghost fields that need to be used for the NS sector. }.
\bea
|\xi(k),k \ra &\equiv &  \xi^{\alpha}(k)p_{\alpha, -1/2} |\sigma, k \ra~, \cr
Q_B |\xi(k), k\ra = 0 &\Rightarrow & k^{\mu} \lt(\gb_{\mu} \xi(k)
\rt)_{\alpha} =0~, \cr
L_0 |\xi(k), k\ra = 0 &\Rightarrow & k^2 = 0~.
\eea
One can further proceed in the similar way. But to count all the
states in the BRST cohomology one may proceed to follow the argument
of Berkovits given in \cite{cohomology}. We shall not repeat the
details here. But one can argue that the analysis goes through for the
NS sector simply by considering all the $SO(8)$ vectors to be periodic
and all the $SO(8)$ spinors to be anti-periodic on the
world-sheet. This includes all the fields appearing in the analysis of
\cite{cohomology} including the infinite number of ghosts for
ghosts. For example the vector field in eqs.(4.3) of \cite{cohomology}
will have the same mode expansion, but the spinor field, in the
present case, will be expanded in terms of half-integer
modes. Having constructed all the transverse creation modes, all the
elements in the BRST cohomology can be obtained simply by applying
those operators freely with a factor of\footnote{In our notation the
  mass-shell condition for the anti-periodic sector reads: $M^2 =
  -{k^2 \over 4}= {1\over 4} (2k^+k^-- \vec k^2)= {1\over 2}(N-{1\over
    2})$. }
$e^{-ik^-X^+}$, with $k^-={N\over k^+}$ on the unique ground state
$|\sigma, k\ra$ with $k^+\neq 0$ (fixed),
$k^-=0$ and $\vec k^2 = 1$. Here $N$ is the total level of
all the creation operators. Notice that $|\sigma, k\ra$ is the NS-sector
analogue of the state in eq.(4.8) of \cite{cohomology} and it does not
have the $c$-ghost dependent additive part. This is because $|\sigma,
k\ra$ is annihilated by $Q_B$ even without the pure spinor constraint
as can be checked by using eqs.(\ref{sigma-def}) and (\ref{QB-modes}).

Although the arguments given in \cite{cohomology} applied to the NS
sector establish the fact that the pure spinor BRST cohomology is isomorphic
to the LCGS Hilbert space \cite{mukhopadhyay04}, we do not have an
explicit construction of this Hilbert space in terms of the pure
spinor degrees of freedom. In the next section we shall achieve this
by performing the analogue of DDF construction in pure spinor
formalism without introducing the infinite number of ghosts for
ghosts.

\sectiono{The DDF Construction}
\label{s:DDF}

Here we shall first construct the DDF states for the periodic
sector. The same for the anti-periodic sector, which will be discussed next,
can be constructed more easily.

\subsectiono{Periodic Sector}
\label{ss:periodic}

The construction is done using the massless vertex operators both in
unintegrated and integrated forms as discussed in appendix \ref{a:massless}\footnote{See \cite{closed-vertex} for a detailed discussion on closed string vertex operators.}.
Unlike in the NSR formalism \cite{GSW}, we exploit the manifest
supersymmetry in the present case to build the whole construction. All
the DDF states will be ghost number one physical states with manifest
$SO(8)$ covariance and will be in one-to-one correspondence with the
states in LCGS formalism. The construction goes through the
following two steps:

\begin{enumerate}
\item
Construct the $SO(8)$ covariant massless DDF states by using the
BRST and supersymmetry properties of the gluon and gluino vertex
operators in the unintegrated form. These operators have dimension
zero and ghost number one. The conjugate states, with respect to which
the massless DDF states form an orthonormal basis, are constructed
using the unintegrated vertex operators of ghost number two states
in the BRST cohomology.

\item
Construct the BRST invariant DDF operators that are in one-to-one
correspondence with the bosonic and fermionic oscillators in LCGS
formalism by using the dimension one, ghost number zero gluon and
gluino vertex operators in the integrated form. One obtains the
physical Hilbert space $\CH_{DDF}$ by applying the creation DDF modes
on the massless states constructed in the first step. Similarly the
conjugate states are obtained by applying the DDF operators on the
massless conjugate states.
\end{enumerate}
In order for the above construction to go through one needs to show
that the DDF operators do have the same algebra of the corresponding LCGS
oscillators. We shall argue that this condition is indeed satisfied
within $\CH_{DDF}$ by proving that the DDF states constructed above
form an orthogonal basis in $\CH_{DDF}$.

Let us now proceed to perform the first step. We shall consider the
massless unintegrated vertex operator in (\ref{ukz}) with the gauge
choice,
\bea
a^+(k)=0~,
\label{a+}
\eea
and the choice of momentum: $k^+(\neq 0)$ and $k^I$ left arbitrary and
$k^-={\vec k^2\over 2k^+}$. According to the on-shell conditions in
(\ref{on-shell}), this implies,
\bea
a^-(k)={1\over k^+} k^I a^I(k)~, \quad
\xi_L(k)= -{1\over \sqrt{2}k^+} k^I \sigma^I\xi_R(k)~.
\label{a-xiL}
\eea
To simplify notations we have depicted the above quantities as
functions of $k$, whereas they are actually functions of only $k^+$ and
$\vec k$. We shall follow the same notation below as well. Using the
above conditions in eq.(\ref{ukz}) one gets,
\bea
u(k,z) &=& a^I(k) v^I(k,z) + \xi_R^{\dot a}(k) s_R^{\dot a}(k,z) ~,\cr
v^I(k,z) &=& b^I(k,z) -{k^I\over k^+} b^+(k,z)~, \cr
s^{\dot a}(k,z) &=& f_R^{\dot a}(k,z) -{k^I\over \sqrt{2}k^+}
\sigma^I_{a\dot a}f_L^a(k,z)~,
\label{massless-unint-lc}
\eea
where $a^I(k)$ and $\xi_R^{\dot a}(k)$ are the independent
components. Therefore the states,
\bea
|I,k\ra \equiv v^I(k, 0) |0\ra~, \quad |\dot a, k\ra \equiv
\sqrt{k^+\over i\sqrt{2}}~s^{\dot a}(k, 0) |0\ra~,
\label{states-I-adot}
\eea
are the physical ground states in $8_v$ and $8_c$ of $SO(8)$. The
particular normalisation chosen will be explained later. Let us now
discuss the supersymmetry transformations of these states. Using
eqs.(\ref{susy-unint}) one finds that in terms of
the $SO(8)$ notations these are given by (up to BRST exact terms),
\bea
Q_L^a |I,k\ra &=& \displaystyle \sqrt{k^+ \over i\sqrt{2}}
\sigma^I_{a\dot a} |\dot a, k\ra~, \cr
Q_L^a |\dot a, k\ra &=& \displaystyle \sqrt{k^+\over i\sqrt{2}}
\sigma^I_{a\dot a} |I,k\ra~, \cr
Q_R^{\dot a} |I,k\ra &=& \displaystyle {1\over
\sqrt{i2\sqrt{2}k^+}}k^J(\delta^{JI}\delta_{\dot a\dot b} +\bar
\sigma^{JI}_{\dot a\dot b}) |\dot b,k\ra~, \cr
Q_R^{\dot a} |\dot b, k\ra &=& \displaystyle {1\over
\sqrt{i2\sqrt{2}k^+}} k^I(\delta^{IJ}\delta_{\dot a\dot b} +\bar
\sigma^{IJ}_{\dot a\dot b}) |J,k\ra~.
\label{susy-Iadot}
\eea
To show the last equality one uses the fact that $k_{\mu}b^{\mu}(k,z)$
is BRST exact which is responsible for the gauge invariance in
(\ref{on-shell}). The above equations are precisely the supersymmetry
transformations of the massless states in LCGS formalism \cite{GSW}.
One can also define a set of conjugate states $\la
I, k|$ and $\la\dot a, k|$ with the following inner products:
\bea
\la I, k|J, l\ra = \delta^{IJ} \delta(k^++l^+) \delta^8(\vec k + \vec
l)~, \quad \la\dot a, k^+|\dot b, l^+\ra = \delta^{\dot a\dot b}
\delta(k^++l^+)\delta^8(\vec k + \vec l)~.
\label{massless-inner-prod}
\eea
These states can be explicitly constructed in terms of ghost number
two zero-mode operators once the $\theta$-expansions of the states
$|I,k\ra$ and $|\dot a, k\ra$ are known. These states are also
annihilated by $Q_B$ and form the BRST cohomology at ghost number
two.

We shall now proceed to the second step where the DDF operators will
be constructed using the vertex operators $\B^{\mu}(k,z)$ and
$\F_{\al}(k,z)$ (see appendix \ref{a:massless}). Let us first
consider the gluon
vertex operator ${\cal B}^I(k^-, z)$ along the light-cone transverse
direction $\mu = I$ with momentum: $k^-\neq 0, k^+=k^I=0$. Using
eqs.(\ref{susy-int}) one shows that only the left moving gluino is
generated under the space-time supersymmetry transformations of this
operator.
\bea
\lt[ Q_L^a, {\cal B}^I(k^-, z) \rt] = 0, \quad \lt[Q_R^{\dot a}, {\cal
B}^I(k^-, z) \rt] = -{i\over \sqrt{2}} k^- \sigma^I_{a\dot a} {\cal
F}_L^a(k^-, z)~,
\label{susy-BI}
\eea
The supersymmetry transformations of ${\cal F}_L(k^-, z)$ are
given by (up to total derivative terms),
\bea
\lt\{Q^a_L, {\cal F}^b_L(k^-, z) \rt\} = \sqrt{2}
\delta^{ab} \B^+(k^-,z)~, \quad
\lt\{Q_R^{\dot a}, {\cal F}^a_L(k^-, z) \rt\} = \sigma^I_{a\dot
a}{\cal B}^I(k^-, z)~.
\label{susy-Fal}
\eea
Notice that $\B^+(0,z)=\partial X^+(z)$ (eq.(\ref{Bmu-Fal-zero-mom}))
and for non-zero $k^-$, $\B^+(k^-,z)$ should be a total derivative as
$k_{\mu}\B^{\mu}(k,z)$ is so, which is responsible for the gauge
invariance in (\ref{on-shell}). Therefore only the first term in the
first equation in (\ref{Bmu-Fal}) will survive at the lowest $\th$-level.
Moreover, the higher order $\th$-terms can not contribute as the operator
has to be a dimension one total derivative. Therefore we should have:
\bea
\B^+(k^-,z)={i\over k^-} \partial e^{-ik^-X^+}(z)~, \quad \quad
k^-\neq 0~.
\label{B+k-}
\eea
Next we define the DDF operators,
\bea
A^I_n(k_0) &\equiv & \oint {dz\over 2\pi } {\cal B}^I(k^-=-nk_0, z)~, \cr
S^a_n(k_0) &\equiv & {1\over \sqrt{-i\sqrt{2}\alpha_0^+}}
\oint {dz \over 2\pi i} {\cal F}^a_L(k^-=-nk_0, z)~,
\label{AInSan}
\eea
where $k_0$ is a real number and $\alpha_0^+$ follows from the
definition given below eq.(\ref{QB-modes}).
All the DDF states are going to
have a fixed
value of $\alpha_0^+$ which is same as that of the states in
(\ref{states-I-adot}). This is simply because none of the above DDF operators
excites momentum along this direction. Also we shall argue in appendix
\ref{a:theorem} that $\partial X^-$ does not appear in any of the DDF
operators (see statement (\ref{result1})), so that $\alpha_0^+$ appears
to be only a c-number in the string of DDF operators in a given DDF state.
 Being constructed out of dimension one primaries, the DDF operators
 commute with all the Virasoro generators. In particular, commuting
 with $L_0$ implies that the action of the $n$-th mode changes the
 level of a state by $-n$. Using eqs. (\ref{BRST-Bmu-Fal}) one can
 argue that while acting on any state of momentum $q$, the operators
 $A^I_n(1/q^+)$ and $S^a_n(1/q^+)$ are BRST invariant.
\bea
\lt[ Q_B, A^I_n(1/q^+)\rt] = 0~, \quad \lt\{ Q_B, S^a_n(1/q^+)\rt\} = 0~.
\label{BRST-inv}
\eea
The supersymmetry transformations of the DDF operators take the
following form on any such state,
\bea
\begin{array}{ll}
\lt[Q_L^a, A^I_n(1/q^+)\rt] = 0 ~, & \lt[Q_R^{\dot a},A^I_n(1/q^+)\rt]
= -{n\over \sqrt{i\sqrt{2}q^+}} S^a_n(1/q^+)~, \\
\lt\{Q_L^a,S_n^b(1/q^+)\rt\} = \sqrt{-i\sqrt{2}q^+} \delta^{ab}\delta_{n,0}~,&
\lt\{Q_R^{\dot a},S^a_n(1/q^+)\rt\} = {1\over
\sqrt{i\sqrt{2}q^+}}\sigma^I_{a\dot a}A^I_n(1/q^+)~.
\end{array}
\label{osc-susy}
\eea
Although the above transformations are precisely the ones expected for
the LCGS oscillators, the commutation relations among the DDF
operators do not quite form the desired algebra because of the terms
higher order in $\th$. Using various OPE's in appendix
\ref{a:notation} one can show:
\bea
\hbox{Res}_{z\to w} {\cal B}^I(k^-, z) {\cal B}^J(p^-, w) &=& ik^-
\delta^{IJ} \partial X^+(w) e^{-i(k^-+p^-)X^+}(w) + {\cal O}(\theta^2)
~,\cr
\hbox{Res}_{z\to w} {\cal F}^a_L(k^-, z) {\cal F}^b_L(p^-, w) &=&
\sqrt{2} \delta^{ab} \partial X^+(w) e^{-i(k^-+p^-)X^+}(w) + {\cal
O}(\theta^2)~, \cr
\hbox{Res}_{z\to w} \B^I(k^-,z) \F_L^a(p^-,w) &=& 0+ {\cal O}(\theta)~,
\label{ResBIFal}
\eea
which imply the following commutation relations,
\bea
\lt[A^I_m(1/q^+), A^J_n(1/q^+) \rt] &=& m \delta^{IJ} \delta_{m,-n} +
{\cal O}(\theta^2)~, \cr
\lt\{S^a_m(1/q^+), S^b_n(1/q^+) \rt\} &=& \delta^{ab} \delta_{m,-n} +
{\cal O}(\theta^2)~, \cr
\lt[A^I_m(1/q^+), S^a_n(1/q^+)\rt] &=& 0+{\cal O}(\theta)~.
\label{osc-alg}
\eea
We shall argue later that these higher order terms will drop off in
the physical Hilbert space that we are going to define. We first
define the following excited states,
\bea
\lt. \begin{array}{l}
|\{(I_i, n_i)\}, \{(a_j, m_j)\}, I, q \ra \\ \\
|\{(I_i, n_i)\}, \{(a_j, m_j)\}, \dot a, q \ra
\end{array} \rt\} \propto \prod_i A^{I_i}_{-n_i}(1/q_0^+) \prod_j
S^{a_j}_{-m_j}(1/q_0^+)
\lt\{\begin{array}{l}
|I, q_0\ra~, \\ \\
|\dot a, q_0\ra ~,
\end{array} \rt.
\label{DDF-states}
\eea
where all the integers $\{n_i\}$ and $\{m_j\}$ are positive definite
and the net momentum $q$ is given by: $q^+=q^+_0, q^-={N\over q_0^+} +
q_0^-, q^I=q_0^I$, with $q_0^2=0$.  $N$ is the total level:
$N=\sum_i n_i+ \sum_j m_j$. The open string mass is given by:
$M^2= -{q^2\over 4}={N\over 2}$.
Therefore all the above DDF
states are annihilated by both $Q_B$ and $L_0$, hence are physical.
They are also
in one-to-one correspondence with the states in LCGS
formalism. In defining the states in eqs.(\ref{DDF-states}) one
follows a particular canonical ordering of all the operators. Although
according to the commutation relations (\ref{osc-alg}) the states with
different ordering are in general completely different states, we
shall later see that effectively they will differ at most by an
overall sign due to the reordering of the fermionic operators.
We define a Hilbert space ${\cal H}_{DDF}$ by the space spanned by the
basis states (\ref{DDF-states}). Next we define the conjugate basis
states,
\bea
\lt. \begin{array}{l}
\la \{(I_i, n_i)\}, \{(a_j, m_j)\}, I, q | \\ \\
\la \{(I_i, n_i)\}, \{(a_j, m_j)\}, \dot a, q |
\end{array} \rt\} \propto \lt. \begin{array}{l}
\la I, q_0| \\ \\
\la \dot a, q_0 | \end{array} \rt \}
\prod_j S^{a_j}_{m_j}(1/q_0^+) \prod_i A^{I_i}_{n_i}(1/q_0^+)~.
\label{conj-DDF-states}
\eea
Similar remarks about the ordering of the operators are in order in
this case as well. If we can now argue that the basis defined this
way is orthogonal such that by choosing the normalisation suitably the
nonzero inner products can be written as,
\bea
\la \{(I_i, n_i)\}, \{(a_j, m_j)\}, I, q |\{(I^{\prime}_i,
n^{\prime}_i)\}, \{(a^{\prime}_j, m^{\prime}_j)\}, I^{\prime},
q^{\prime} \ra &=& \delta_{\{I_i,n_i\}, \{I^{\prime}_i,
  n^{\prime}_i\}} \delta_{\{a_j,m_j\}, \{a^{\prime}_j, m^{\prime}_j\}}
\cr
&& \delta_{I,I^{\prime}} \delta(q^++q^{\prime +}) \delta(\vec q+ \vec q^{\prime})~,\cr
\la \{(I_i, n_i)\}, \{(a_j, m_j)\}, \dot a, q |\{(I^{\prime}_i,
n^{\prime}_i)\}, \{(a^{\prime}_j, m^{\prime}_j)\}, \dot a^{\prime},
q^{\prime} \ra &=& \delta_{\{I_i,n_i\}, \{I^{\prime}_i,
  n^{\prime}_i\}} \delta_{\{a_j,m_j\}, \{a^{\prime}_j,
  m^{\prime}_j\}}\cr
&&\delta_{\dot a, \dot a^{\prime}}\delta(q^++q^{\prime +}) 
\delta(\vec q + \vec q^{\prime})~, \cr &&
\eea
then it would imply that the DDF operators have the desired algebra in
${\cal H}_{DDF}$.
\bea
\lt[A^I_m(1/q_0^+), A^J_n(1/q_0^+) \rt]_{{\cal H}_{DDF}} &=&
\delta^{IJ} \delta_{m,-n}~, \cr
\lt\{S^a_m(1/q_0^+), S^b_n(1/q_0^+) \rt\}_{{\cal H}_{DDF}} &=&
\delta^{ab} \delta_{m,-n}~, \cr
\lt[A^I_m(1/q_0^+), S^a_n(1/q_0^+)\rt]_{{\cal H}_{DDF}} &=& 0 ~.
\label{osc-alg-HDDF}
\eea
Certainly the validity of our construction crucially relies on the
orthogonality of the DDF states defined in this section. We shall
prove it section \ref{ss:validity}.

Before going into the DDF construction for the anti-periodic sector
let us explain the normalisation chosen in eqs.(\ref{states-I-adot}).
Using the second equation in each of (\ref{AInSan}) and
(\ref{Bmu-Fal-zero-mom}) one shows,
\bea
S^a_0 = {1\over \sqrt{-i\sqrt{2}\alpha_0^+}} Q^a_L~, \quad \quad
\lt\{ S^a_0, S^b_0\rt\} = \delta^{ab}~,
\label{S0}
\eea
which are exact in $\th$ expansions and as desired for the fermionic
zero modes in LCGS formalism \cite{GSW}. Then using the first two
equations in (\ref{susy-Iadot}) one shows,
\bea
S^a_0 \pmatrix{|I,q\ra \cr |\dot a, q\ra }= {1\over \sqrt{2}}
\sigma^I_{a\dot a}
\pmatrix{|\dot a, q\ra \cr |I, q\ra }~,
\label{S0action}
\eea
which is also expected.

\subsectiono{Anti-periodic Sector}
\label{ss:anti-periodic}

Let us now turn to the construction of DDF states in the anti-periodic
sector. Supersymmetry, which has played a crucial role in such
construction in the periodic sector, is broken in this
case. Nevertheless we shall now see that the non-supersymmetric open
string spectrum can be obtained by a simple generalisation of the
previous construction. In the anti-periodic sector any space-time
fermion has half-integer modes whereas all the space-time bosons have
the same integer mods as in the previous case. This means that the
construction of the bosonic oscillators $A^I_n(1/q^+)$ still goes
through with the commutation relation as given in
(\ref{osc-alg-HDDF}). We define the half-integer fermionic modes in
the following way,
\bea
S^a_r(1/q^+) = {1\over \sqrt{-i\sqrt{2} \alpha_0^+}} \oint {dz\over
  2\pi i} \F^a_L(k^-=-r/q^+, z)~,
\label{Sar}
\eea
where $r\in \Z +1/2$. In the present sector the vacuum $|\sigma \ra$
is twisted so
that it produces branch cut for the space-time fermions. The action of
$S^a_r(1/q^+)$ is well defined on any excited sate of momentum $q$ on
this vacuum, because the branch cut in the space-time fermions is
cancelled by the branch cut produced by the half-integer units of
momentum in the DDF vertex $\F^a_L(k^-=-r/q^+, z)$. Using the second
equation in (\ref{BRST-Bmu-Fal}) and the anti-periodicity of the
fermions one can argue that these operators are BRST exact on any
state of momentum $q$,
\bea
\lt\{Q_B, S^a_r(1/q^+)\rt\} =0~.
\eea
The physical Hilbert space ${\cal H}_{DDF}$ is now defined to be
expanded by the following ghost number one states,
\bea
|\{(I_i, n_i)\}, \{(a_j, r_j)\}, q \ra \propto \prod_i
A^{I_i}_{-n_i}(1/q_0^+) \prod_j S^{a_j}_{-r_j}(1/q_0^+) |\sigma,
q_0\ra~,
\eea
with fixed $q_0^+$ such that $q_0^2=1$. The net
momentum $q$ is now given by, $q^+=q_0^+$, $q^-={N\over q_0^+} +
q_0^-$, $q^I=q_0^I$, where the net level is:
$N=\sum_i n_i + \sum_j r_j$. The open string mass is now given by:
$M^2 = {1\over 2}\lt(N-{1\over 2}\rt)$.
Proceeding similarly as in the periodic case by
defining the conjugate states one argues the following commutation
relations in the physical Hilbert space,
\bea
\lt\{S^a_r(1/q^+), S^b_s(1/q^+)\rt\}_{{\cal H}_{DDF}} = \delta^{ab}
\delta_{r,-s}~,\quad
\lt[A^I_n(1/q^+), S^a_r(1/q^+) \rt]_{{\cal H}_{DDF}} = 0~.
\label{osc-alg-HDDF-ant-per}
\eea
Again one needs the DDF basis to be orthogonal, an issue that will be
discussed in the next section.

\subsectiono{Validity of the Construction}
\label{ss:validity}

Validity of the DDF construction as done above relies on the
orthogonality of the DDF states both in the periodic and anti-periodic
sectors. Here this orthogonality will be proved. We shall, for
definiteness, consider the periodic sector, generalisation to the
anti-periodic sector being obvious.

We first notice, with the relation (\ref{S0action}) in mind, that an
arbitrary inner product between that states in (\ref{DDF-states}) and
(\ref{conj-DDF-states}) can be given the following form,
\bea
{\cal I} &=& \la J|\prod_i S^{b_i}_{\bar m_i} \prod_j A^{J_j}_{\bar
n_j} \prod_k A^{I_k}_{-n_k} \prod_l S^{a_l}_{-m_l} |I\ra~,
\label{calI}
\eea
if we allow the zero modes for the fermionic oscillators. To reduce
clutter, we have suppressed the momentum specification. The indices
$i,j,k,l$ run up to arbitrary positive integers. The bosonic mode
numbers $\bar n_j, n_k$ are positive definite while the fermionic ones
$\bar m_j, m_l$ are positive, including zero. Next we notice that, due
to the $\th$-charge conservation, the matrix element of any operator
of nonzero $\th$-charge between two vector ground states is zero. The
only non-trivial operators that can have non-zero matrix elements are
rotation generators which have zero $\th$-charge. Therefore after
expanding all the operators in (\ref{calI}) in powers of $\th$ the
only terms that give non-zero results are those for which sum of the
$\th$-charges of all
the operators in a given product is zero. Here is a special example of
a term that is potentially non-zero:
\bea
&& {\cal I}_{special}=\la J|\prod_i S^{(\bar o_i) b_i}_{\bar m_i}
\prod_j A^{(0)J_j}_{\bar n_j} \prod_k A^{(0)I_k}_{-n_k}
\prod_l S^{(o_l)a_l}_{-m_l} |I\ra~, \cr
&& \bar o_i, o_l = \pm 1~, \quad (\sum_i\bar o_i + \sum_l o_l)=0~,
\label{calIspecial}
\eea
where the extra index in the parenthesis refers to the $\th$-charge of
the term in the $\th$-expansion of the relevant operator. We shall see
later in what sense this inner product is special. Let us first try to
compute the inner product using the following commutation relations,
\bea
\lt[A^{(0)I}_m(1/q_0^+), A^{(0)J}_n(1/q_0^+) \rt] &=&
m \delta^{IJ} \delta_{m,-n}~, \cr
\lt[A^{(0)I}_m(1/q_0^+), S^{(-1)a}_n(1/q_0^+) \rt] &=&0~, \cr
\lt\{S^{(-1)a}_m(1/q_0^+), S^{(-1)b}_n(1/q_0^+) \rt\} &=&0 ~, \cr
\lt\{S^{(-1)a}_m(1/q_0^+), S^{(1)b}_n(1/q_0^+) \rt\} &=&{1\over 2}
\delta^{ab} \delta_{m,-n}~.
\label{additional-comm}
\eea
The commutation relations involving $S^{(-1)}$'s (to simplify notation
we are suppressing the indices that are not relevant for our
discussion) guarantee that in order for the inner product to be
non-zero we should have the following condition satisfied: let $n^+$
and $n^-$ be the number of positively modded fermionic operators with
$\th$-charge $+1$ and $-1$ respectively, then the number of negatively
modded fermionic operators with $\th$-charge $+1$ and $-1$ are given by
$n^-$ and $n^+$ respectively. One can then move $S^{(-1)}$'s towards
right or left, as
appropriate, to absorb all the $S^{(1)}$'s. This way one gets rid of
all the fermionic operators. Then the inner product of the bosonic
operators can easily
be found by using the first equation in
(\ref{additional-comm}). Therefore the final result should be,
\bea
{\cal I}_{special} \propto \delta_{\{J_j,\bar n_j\}, \{I_k,\bar n_k\}}
\delta_{\{b_i,\bar m_i\}, \{a_l,m_l\}} \delta_{IJ}~.
\label{calIspecial-result}
\eea
Certainly there is an obvious delta function involving momenta, which
is suppressed in the above expression. One does not have other
symmetry terms originating from interchange of the operators as the
basis states have been defined with an ordering.

We shall now argue that the only terms that are non-zero in
${\cal I}$ are of the type ${\cal I}_{special}$.
${\cal I}_{special}$ is the kind of terms in ${\cal I}$ that come with
the minimum number of $S^{(-1)}$'s. All the other terms with zero
total $\th$-charge can be obtained by replacing some of the operators
(both bosonic and fermionic) in ${\cal I}_{special}$ by the
corresponding ones with higher $\th$-charge and balancing the total
$\th$-charge by adding suitable number of extra
$S^{(-1)}$'s. More we bring in
higher $\th$-charge operators, bigger we make the mismatch between the
numbers of $S^{(-1)}$'s and $S^{(1)}$'s. If the higher $\th$-charge
operators commute with $S^{(-1)}$'s then the result will be zero. But
generically this will not be the case. The final result can still be
zero if the collection of all the higher $\th$-charge operators
are unable to absorb all the extra $S^{(-1)}$'s through
commutators. The necessary and sufficient condition for this to happen
is the fact that the term in either of the DDF vertex operators
${\cal B}^I(k^-, z)$ and ${\cal F}_L^a(k^-,z)$ at the $n$-th order in
 $\th$-expansion does not have a term with charge $(n,0)$ for $n>1$.
According to our notation, a term with charge $(p,q)$ has
left ($\th_L$) and right ($\th_R$) moving $\th$-charges $p$ and $q$
respectively. The above requirement is satisfied due to a theorem
that we call {\it absence of maximal left moving $\th$-charges}
which is stated and proved in appendix \ref{a:theorem}. This establishes
the fact that the inner product in
eq.(\ref{calI}) is proportional to the right hand side of
(\ref{calIspecial-result}) and therefore the DDF basis states are
orthogonal.

\sectiono{Physical Components of Boundary States for Instantonic D-branes}
\label{s:physical}

A particular approach of studying open string boundary conditions and
D-brane boundary states have been considered in \cite{mukhopadhyay05}
where one writes down the boundary conditions and boundary states in
the free CFT by relaxing the pure spinor constraint. These boundary
conditions and boundary states are easy to construct and are in
one-to-one correspondence with the actual boundary conditions and
boundary states of the constrained CFT. The boundary conditions in the
free CFT produce the correct reflection property between the
holomorphic and anti-holomorphic parts of any bulk insertion that is
allowed in the pure spinor CFT. With suitable choice of vertex
operators the boundary states are expected to produce correct results
for all the closed string one-point functions of the actual
theory. But these boundary states, as one might already expect, are
not suitable for computation of the cylinder diagram. The reason is
two-fold which we list below:
\begin{enumerate}
\item
Having been constructed in a bigger Hilbert space, these boundary
states contain degrees of freedom which do not belong to the actual
theory. Let us call them {\it unphysical} degrees of freedom.
\item
The boundary states have been constructed in the gauge unfixed theory.
\end{enumerate}
The first problem could be solved simply by throwing away all the {\it
  unphysical} degrees of freedom. A covariant pure spinor boundary
state at ghost number $(1,1)$ can be defined in the following way:
\bea
|\hbox{B}\ra_{PS} = \sum_{i_{PS}} \varphi^{(B)}_{i_{PS}} |i_{PS}\ra~,
\label{BPS}
\eea
where $\lt\{|i_{PS}\ra \rt\}$ is a complete basis of ghost number
$(1,1)$ states in pure spinor formalism (with the constraint
imposed). The one point functions
$\varphi^{(B)}_{i_{PS}}$ can be computed following the prescription of
\cite{mukhopadhyay05} using the boundary state $|\hbox{B}\ra_{free}$
constructed in the free CFT. One might think that the projected
boundary state $|\hbox{B}\ra_{PS}$ could be evolved by world-sheet
time evolution to compute the cylinder diagram. As argued and
demonstrated explicitly through the long range force computation in
\cite{mukhopadhyay05}, this is not true as the boundary sate still
includes gauge degrees of freedom. In NSR formalism these gauge
degrees of freedom are removed by a simple gauge fixation (Siegel
gauge). It is in this particular gauge the closed string propagator in
Schwinger parametrisation has an interpretation of world-sheet time
evolution. It is not clear how to achieve this in pure spinor
formalism. In summary, the problem is to find a suitable further
projection of the boundary state $|B\ra_{PS}$ to remove the gauge
degrees of freedom so that the projected boundary state can be evolved
by the world-sheet time evolution. Here we shall achieve this by
projecting the boundary states onto the physical Hilbert space ${\cal
  H}_{DDF}$ constructed in the closed string sector. Since the DDF
construction gives explicit expression for the LCGS variables
in terms of the pure spinor variables, one
would expect that the projected boundary states would take the same
form of the LCGS boundary states in terms of the DDF
operators. We shall see that this expectation is actually correct.
Before going into the further details, we should mention that because
of the special kinematical condition that $q_0^+$ is fixed and
non-zero for all the DDF states, these are suitable to extract the
physical components of only the instantonic boundary states which
impose Dirichlet boundary conditions along both the light-cone
directions. For a lorentzian D-brane having Neumann boundary condition
along both the light-cone directions we need states which have both
$q_0^{\pm}$ to be zero. For D-branes which have Neumann boundary
condition along one of the light-cone directions and Dirichlet along
the other $q_0^+$ needs to vary over the allowed states. We have
discussed in appendix (\ref{a:Dinstanton}) the boundary conditions
for the instantonic D-branes in the free CFT,
following the same approach of \cite{mukhopadhyay05}.

The DDF states in the closed string theory can be constructed simply
by constructing the DDF operators $A^I_n(1/q_0^+)$,
$S^a_n(1/q_0^+)$ and $\tilde A^I_n(1/q_0^+)$, $\tilde S^a_n(1/q_0^+)$,
as in the previous section, in the left and right moving sectors
separately. Then we define, as before, the DDF states which may be
denoted as,
\bea
\pmatrix{|\{I_i,n_i\}, \{a_j,m_j\}, I, q\ra_L \cr
|\{I_i,n_i\}, \{a_j,m_j\},\dot a ,q\ra_L} \otimes
\pmatrix{|\{\tilde I_i,\tilde n_i\}, \{\tilde a_j,\tilde m_j\}, \tilde
I, q\ra_R \cr
|\{\tilde I_i,\tilde n_i\}, \{\tilde a_j,\tilde m_j\}, \tilde{\dot
a},q \ra_R}~,
\label{DDF-states-closed}
\eea
with $N=\tilde N$, which comes, as usual, from the $L_0=\tilde L_0$
constraint. The above states correspond to all the physical degrees of
freedom and are on-shell with mass given by: $M^2 = 2N$.
Given the ghost number $(1,1)$ covariant boundary state
$|\hbox{Inst}\ra_{PS}$ of an instantonic D-brane in pure spinor
formalism, its physical component is given by,
\bea
|\hbox{Inst},q^+ \ra_{phys} = \sum_i |i, q \ra \la i, q |
\hbox{Inst} \ra_{PS} = \sum_i \varphi^{(Inst)}_i |i, q\ra ~,
\label{Inst-phys}
\eea
where the states $|i, q\ra$ are the ghost number $(1,1)$ orthonormal
basis states in ${\cal H}_{DDF}$ given in (\ref{DDF-states-closed})
with a fixed $q^+$ and the states $\la i, q^+|$ are the corresponding
conjugate states with ghost number $(2,2)$. The sum over $i$ in the
above equation includes integration over spatial components of
momentum as well as the discrete levels. The coefficients
$\varphi^{(Inst)}_i$ can be computed using the boundary states
$|\hbox{Inst}\ra_{free}$ constructed in the free theory following the
prescription of \cite{mukhopadhyay05}. Therefore given
$|\hbox{Inst}\ra_{free}$, $|\hbox{Inst}, q^+ \ra_{phys}$ can be
constructed unambiguously. But to get a closed form expression for
$|\hbox{Inst}, q^+ \ra_{phys}$ we shall proceed in a less direct
way. $SO(8)$ covariant boundary states for the BPS and non-BPS
instantonic D-branes in LCGS formalism are already known
\cite{green96, nemani, mukhopadhyay04}. The pure spinor boundary
state in (\ref{Inst-phys}) is expected to take the same form as the
corresponding one in LCGS formalism. We shall argue that this
is in fact true by deriving the gluing conditions satisfied by the DDF
operators on $|\hbox{Inst}, q^+\ra_{phys}$  and showing that they are
same as the corresponding ones in LCGS formalism.

The DDF gluing conditions will be obtained by using boundary
conditions written in open string channel and then converting that to
the closed string channel as needed. Using the mode expansion of $X^+$
one can argue that in the closed string channel at $\tau =0$,
\bea
\lt[ e^{ink_0X^+_L}(e^{i\sigma}) - e^{-ink_0X^+_R}(e^{-i\sigma}) \rt]
|\hbox{Inst}, q^+\ra_{phys} =0~,
\label{bc-exp-X+}
\eea
which can be used, in addition to the boundary condition in
eq.(\ref{UVbc-BPS}), to argue that for an instantonic BPS D$p$-brane
one has,
\bea
\lt. \begin{array}{r}
\lt[\B^I(-nk_0, e^{i\sigma}) - e^{-2i\sigma} ({\cal M}^V)^I_{~J}
\tilde \B^J(nk_0, e^{-i\sigma}) \rt] \\ \\
\lt[\F_L^a(-nk_0, e^{i\sigma}) +i\eta e^{-2i\sigma} {\cal M}^S_{ab}
  \tilde \F_L^b(nk_0, e^{-i\sigma}) \rt]
\end{array} \rt\}
|\hbox{Inst}_p,\eta, q^+\ra_{phys} &=& 0~,
\label{BIFL-gluing-BPS}
\eea
where ${\cal M}^V$ is the 8-dimensional block of $M^V$ (as defined
below eq.(\ref{Xbc})) corresponding to the light-cone transverse
directions. We define matrices ${\cal M}^S$ and ${\cal M}^C$ in the
following way,
\bea
M^S = \bar M^S = \pmatrix{{\cal M}^S_{ab} & 0 \cr 0 &
{\cal M}^C_{\dot a \dot b}}~,
\eea
where the matrices $M^S$ and $\bar M^S$ are defined in
eqs(\ref{MSMbS}). Upon recalling the definitions (\ref{AInSan}),
eqs.(\ref{BIFL-gluing-BPS}) readily give the following
gluing conditions for the DDF operators.

\bea
\lt.
\begin{array}{r}
\lt[ A^I_n(1/q^+) - ({\cal M}^V)_{IJ} \tilde A^J_{-n}(1/q^+) \rt] \\ \\
\lt[ S^a_n(1/q^+) + i \eta {\cal M}^S_{ab} \tilde S^b_{-n}(1/q^+) \rt]
\end{array} \rt\}
|\hbox{Inst}_p,\eta, q^+\ra_{phys}
= 0~, \quad \quad \forall n \in \Z~.
\label{DDF-gluing-BPS}
\eea

The bosonic part of the gluing conditions satisfied by a non-BPS
D-instanton takes the same form as in (\ref{DDF-gluing-BPS}). To
obtain the fermionic part we proceed as follows: Writing the
integrated gluino vertex operator in the following form:
$\F_{\alpha}(k,z) = {\cal G}_{\alpha}(z) e^{ik.X}(z)$, we first use
covariance to argue that ${\cal G}_{\alpha}(z)$ satisfies the same
boundary condition as $p_{\alpha}(z)$ as given by the last equation in
(\ref{bcnBPS}). The $SO(8)$ decomposition of this boundary condition
gives on UHP,
\bea
{\cal G}_L^a(z) {\cal G}_L^b(w) = -{\cal M}^{ab}_{cd} ~\tilde {\cal
  G}_L^c(\zb) \tilde {\cal G}_L^d(\wb) ~, \quad \quad \hbox{at }
z=\zb~, w=\wb~,
\label{bcnBPSG}
\eea
where the coupling matrix ${\cal M}^{ab}_{cd}$ is given by,
\bea
{\cal M}^{ab}_{cd} = {1\over 8} \delta_{ab} \delta_{cd} +
{1\over 16} \sum_{I,J} \lam_I \lam_J \sigma^{IJ}_{ab}
\sigma^{IJ}_{cd} + {1\over 192} \sum_{\{IJKL\}\in {\cal K}^{(4)}}
\lam_I \lam_J \lam_K \lam_L \sigma^{IJKL}_{ab}\sigma^{IJKL}_{cd}~,
\label{Mabcd}
\eea
where we have used $({\cal M}^V)^I_{~J} = \lam_I \delta^{IJ}$ for
notational simplicity and ${\cal K}^{(4)}$ is defined, analogously
to ${\cal K}^{(5)}$ in eqs.(\ref{calMs}), for sets of four integers
instead of five. Using (\ref{bcnBPSG}) and (\ref{bc-exp-X+}) one can
argue that the following condition is satisfied for a non-BPS
instantonic D$p$-brane in the closed string channel at $\tau =0$,
\bea
\lt[\F_L^a(-mk_0, e^{i\sigma}) \F_L^b(-nk_0,e^{i\sigma^{\prime}}) +
e^{-2i(\sigma +\sigma^{\prime})} {\cal M}^{ab}_{cd} \tilde
\F_L^c(mk_0, e^{i\sigma}) \F_L^d(nk_0,e^{i\sigma^{\prime}}) \rt]
|\hbox{Inst}_p ,q^+\ra_{phys} = 0~, \cr
\label{FL-gluing-nBPS}
\eea
which implies the following gluing condition for the fermionic DDF operators,
\bea
\lt[S^a_m(1/q^+) S^b_n(1/q^+) + {\cal M}^{ab}_{cd} \tilde S^c_{-m}(1/q^+)
\tilde S^d_{-n}(1/q^+) \rt] |\hbox{Inst}_p, q^+\ra_{phys} = 0 ~,
\quad \forall m,n \in \Z~. \cr
\label{DDF-gluing-nBPS}
\eea
Eqs.(\ref{DDF-gluing-BPS}) and (\ref{DDF-gluing-nBPS}) are precisely
the same gluing conditions satisfied by the BPS and non-BPS instantonic
D-brane boundary states in LCGS formalism as discussed in
\cite{green96} and \cite{mukhopadhyay04} respectively. The physical
components of the D-instanton boundary states in pure spinor formalism
can therefore be found simply by replacing the LCGS
oscillators by the corresponding DDF operators constructed here in the
expressions for the boundary states found in \cite{green96} and
\cite{mukhopadhyay04} (with the obvious change of notations for the
$SO(8)$ vector and spinor matrices). Notice that the physical
components of the D-instanton boundary  states constructed this way
have very complicated expressions in terms of the pure spinor
variables as the DDF operators have $\theta$-expansions. But for
computations restricted to $\CH_{DDF}$ these states
behave as simply as the boundary states in LCGS formalism.

\sectiono{The Cylinder Diagram}
\label{s:cylinder}

Having removed all the unphysical degrees of freedom which one should
not let propagate in the cylinder diagram we can now evolve the
projected boundary state $|\hbox{Inst}, q^+\ra_{phys}$ by the closed
string propagator $1/(L_0+\tilde L_0)$.
\bea
{\cal C}(X^+,X^-) &\propto & \int dq^+ dq^- \la \hbox{Inst},-q^-,-q^+|
{e^{iq^+X^- + i q^-X^+}\over L_0 + \tilde L_0}
|\hbox{Inst}^{\prime}, q^-, q^+\ra~,
\eea
where $X^{\pm}$ are the separation between the two branes along
the light-cone directions. There
is also a separation in the transverse direction which we have
suppressed. The states have
been allowed arbitrary $q^-$ as required by the Fourier transform of
the position eigen states. But we shall see that the propagating
states will have the on-shell value. Writing $(L_0+ \tilde L_0)$ in
the following form,
\bea
L_0 + \tilde L_0=-2p^+(p^--H)~,
\eea
where,
\bea
H= {1\over 2p^+} (\vec p^2 + N +\tilde N)~,
\eea
with $N=N^{(X)}+N^{(p,\th)}+N^{(w,\lambda)}$ (similarly for the right
moving sector) and following the same steps as in \cite{bergman02}
one arrives at the following expression for the cylinder diagram,
\bea
{\cal C}(X^+,X^-)
&\propto & \int_0^{\infty} {d\tau \over \tau} e^{{iX^+X^-\over 2\pi
    \tau}} \la \hbox{Inst}, -q^+| e^{i\pi \tau(\vec p^2 + N + \tilde
  N)}|\hbox{Inst}^{\prime},q^+\ra~,
\eea
where $\tau = {X^+\over 2\pi q^+}$ can be easily identified with the
modulus of the lorentzian cylinder. Going to the euclidean world-sheet
by the Wick rotation: $\tau \to it$, one arrives at,
\bea
{\cal C}(X^+,X^-)
&\propto & \int_0^{\infty} {dt \over t} e^{{X^+X^-\over 2\pi t}} \la
\hbox{Inst}, -q^+| e^{-\pi t (\vec p^2 + N + \tilde
  N)}|\hbox{Inst}^{\prime},q^+\ra~.
\eea
Computation of this quantity is well-understood and the open-closed
duality is manifest in the result.

\sectiono{Conclusion}
\label{s:conclusion}

The open string boundary conditions and boundary states for both BPS
and non-BPS D-branes in pure spinor formalism were written down in
\cite{mukhopadhyay05} in the unconstrained CFT by relaxing the pure
spinor constraint. It was argued that these boundary conditions and
boundary states are suitable to compute boundary conformal field
theory correlators and closed string one point functions
respectively. But one can not evolve these boundary states according
to the world-sheet time evolution to compute the cylinder
diagram with manifest open-closed duality. This is because one does
not know how to remove the gauge degrees of freedom propagating along
the cylinder. The cylinder diagram can still be computed if one is able to
project these boundary states onto a physical Hilbert space free of
such degrees of freedom.

In this paper we have explicitly constructed such a physical Hilbert space
in pure spinor formalism. By exploiting the manifest supersymmetry of
the formalism, this Hilbert space has been constructed by performing a
supesymmetric version of the usual DDF construction \cite{DDF}.
This gives an explicit realisation of all the states obtained in LCGS
formalism. The validity of our construction has been justified by
proving that the DDF operators constructed here have the same
commutation relations as those of the LCGS oscillators in the physical
Hilbert space. Outside this Hilbert space the commutators have
non-trivial $\th$-expansions.

The DDF construction for open strings on BPS D-branes (closed strings)
implicitly defines the ghost number one (two) unintegrated vertex operators
for all the string states in the BRST cohomology with special kinematical
conditions\footnote{See \cite{berkovits02} for computation of the
covariant vertex
operators at the first massive level.}. All these vertex operators
take a form where the ghost
number of the operator is contributed only by the zero modes of the
pure spinor ghosts. Using the boundary conditions in
\cite{mukhopadhyay05} it is argued that there are two sectors of open
strings on a non-BPS D-brane: periodic (R) and anti-periodic (NS). The
analysis for the R sector goes in the same way as that corresponding
to the BPS D-branes. Although the DDF construction for the NS sector
is well defined, the unintegrated vertex operators, that are needed for
the scattering amplitude computations, can not be derived
from this construction unless the vertex operator for the unique
ground state, which represents the open string tachyon, is
understood\footnote{I thank N. Berkovits for
  discussion on this point.}. 
Understanding of how to
explicitly construct this ground state is an interesting and important
open question. This is an example of a more generic question of how to
construct the ground states for open strings stretching between branes
at angles which will allow more general boundary conditions.

Going back to the discussion of boundary states, we derive the gluing
conditions for the DDF operators satisfied on the boundary states for both
the BPS and non-BPS instantonic D-branes. We show that these
conditions are exactly the same as those satisfied by the oscillators
in LCGS formalism \cite{green96, mukhopadhyay04}. Therefore the
projected boundary states in pure spinor formalism can be obtained
simply by replacing the LCGS oscillators by the DDF operators
constructed here in the expressions for the boundary states written down
in \cite{green96, mukhopadhyay04}. This construction offers an
explicit embedding of all the computations
that can possibly be done in LCGS formalism into pure spinor
formalism, a particular example being the cylinder diagram with manifest
open-closed duality. However, computing the cylinder diagram using a
covariant boundary state still remains an open question. It is
important to identify the relevant covariant basis for which the techniques of
\cite{berkovits05} may prove useful.

\medskip
\centerline{\bf Acknowledgement}
\noindent
I wish to thank S. R. Das, M. B. Green, K. Hashimoto, A. Sinha
and N. Suryanarayana for useful discussion and N. Berkovits for helpful
communications and comments on a preliminary draft.
This work was partially supported by DOE grant DE-FG01-00ER45832 and PPARC.

\appendix

\sectiono{Notation and Convention} \label{a:notation}

We follow the same notation and convention for the 32 and 16
dimensional gamma matrices as given in
\cite{mukhopadhyay05}. Therefore all the gamma matrix
properties and Fiertz identities summarised in the relevant appendix
of \cite{mukhopadhyay05} still hold. Here we shall consider an explicit
$SO(8)$ decomposition. We define the light-cone components $A^{\pm}$ of a
10-dimensional vector $A^{\mu}$ in the following way,
\bea
A^{\pm}= {1\over \sqrt{2}} (A^0\pm A^9)~.
\label{lc-comp}
\eea
A 16-dimensional chiral spinor $\xi$ of either chirality is decomposed
into the left and right moving $SO(8)$ spinors in the following way,
\bea
\xi = (\xi_L^a , ~\xi_R^{\dot a})~.
\eea
The 16-dimensional gamma matrices of \cite{mukhopadhyay05} are given by,
\bea
\g^0 = -\gb^0 = \Iop_{16}~, \quad \g^I=\gb^I=\pmatrix{0&
  \sigma^I_{a\dot a}
\cr \bar \sigma^I_{\dot a a} & 0}~,\quad \g^9=\gb^9= \g^1\g^2\cdots
\g^8 =
\pmatrix{\Iop_8 & 0\cr 0&-\Iop_8}~, \cr
\eea
where $\bar \sigma = (\sigma)^T= \sigma =\sigma^*$.

We shall now collect some of the expressions that are relevant for
open strings and are directly needed for the computation of the
present paper. We work in the $\alpha^{\prime}=2$ unit, such that,
\bea
X^{\mu}(z) X^{\nu}(w) \sim - \eta^{\mu \nu} \log |z-w|~.
\label{X-X-ope}
\eea
The supersymmetry charge is given by,
\bea
Q_{\alpha} &=& \oint {dz\over 2\pi i} q_{\alpha}(z)~, \cr
q_{\alpha} &=& p_{\alpha} + {1\over 2} \lt(\gb^{\mu}\theta \rt)_{\alpha}
\partial X_{\mu} + {1\over 24} \lt(\gb^{\mu}\theta \rt)_{\alpha}
\lt(\theta \gb_{\mu}\partial \theta \rt)~.
\label{susy-charge}
\eea
Using (\ref{X-X-ope}) and,
\bea
p_{\alpha}(z) \th^{\bt}(w) \sim {\delta_{\alpha}^{~\bt}\over z-w}~,
\label{p-th-ope}
\eea
one can derive the following supersymmetry algebra,
\bea
\lt\{ Q_{\al}, Q_{\bt} \rt \} =\gb^{\mu}_{\al \bt} \oint
{dz\over 2\pi  i} \partial X_{\mu}(z)~,
\eea
which takes the following form in the $SO(8)$ notation,
\bea
\lt\{Q_L^a, Q_L^b \rt\} &=& \sqrt{2} \delta^{ab} \oint {dz\over 2\pi i}
\partial X^+(z)~,\cr \lt\{Q_L^a, Q_R^{\dot a} \rt\} &=&
\sigma^I_{a\dot a}
\oint {dz\over 2\pi i} \partial X^I(z)~, \cr
\lt\{Q_R^{\dot a}, Q_R^{\dot b} \rt\} &=& \sqrt{2}
\delta^{\dot a\dot b}  \oint {dz\over 2\pi i} \partial X^-(z)~.
\eea
The BRST charge is given by,
\bea
Q_B &=& \oint {dz\over 2\pi i} q_B(z)~, \cr
q_B&=& \lambda^{\alpha}(z) d_{\alpha}(z)~, \quad
d_{\alpha} = p_{\alpha} - {1\over 2} (\bar \gamma ^{\mu} \th)_{\alpha}
\partial X_{\mu} - {1\over 8} (\bar \gamma^{\mu} \th)_{\alpha} (\th \bar
\gamma_{\mu} \partial \th)~.
\label{QB}
\eea
The fermionic matter and the pure spinor ghost contributions to the Lorentz
currents $M^{\mu \nu}(z)$ and $N^{\mu \nu}(z)$ respectively are given
by,
\bea
M^{\mu \nu} = -{1\over 2} (p\gamma^{\mu \nu}\th)~, \quad N^{\mu \nu} =
{1\over 2} (w\gamma^{\mu \nu}\lambda)~.
\eea
They form $SO(9,1)$ current algebra at levels $4$ and $-3$ respectively
and satisfy the following OPE's,
\bea
M^{\mu \nu}(z) \th^{\alpha}(w) &\sim& {1\over 2(z-w)} \lt( \gamma^{\mu
  \nu} \th(w) \rt)^{\alpha} ~, \quad M^{\mu \nu}(z) p_{\alpha}(w) \sim
{1\over 2(z-w)} \lt(\bar \gamma^{\mu \nu}p(w)\rt)_{\alpha}~,\cr
N^{\mu \nu}(z) \lambda^{\alpha}(w) &\sim& {1\over 2(z-w)} \lt(
\gamma^{\mu \nu} \lambda (w) \rt)^{\alpha} ~.
\eea
Finally we define the Virasoro zero mode in the following way:
\bea
L_0 = {\alpha_0^2\over 2} + N^{(X)} + N^{(p,\th)} + N^{(w,\lambda)} + a~,
\label{L0}
\eea
where $N^{(X)}$,
$N^{(p,\th)}$ and $N^{(w,\lambda)}$ are the level operators for the
bosonic matter, fermionic matter and bosonic ghost sectors
respectively. $N^{(X)}$ and $N^{(p,\th)}$ are defined in the usual
way. For the ghost sector this may be defined through the following
commutation relations:
\bea
[N^{(w,\lambda)}, \lambda^{\alpha}_r] = -r
\lambda^{\alpha}_r~, \quad [N^{(w,\lambda)}, N^{\mu \nu}_n] &=& -n
N^{\mu \nu}_n~, \quad [N^{(w,\lambda)}, J_n] = -n J_n~,
\label{Nwlambda-def}
\eea
where $n$ is an integer and $r$ is an integer or half integer
depending on whether we are considering the periodic or anti-periodic
sector respectively. $N^{\mu \nu}_n$ and $J_n$ are the modes of the currents
$N^{\mu \nu}(z)$ and $J(z) \propto w_{\alpha}(z) \lambda^{\alpha}(z)$
respectively. The normal ordering constant $a$ in eq.(\ref{L0}) is
given by,
\bea
a=\lt\{ \begin{array}{ll}
0~, \quad \quad & \hbox{periodic sector,} \\ & \\
-{1\over 2} ~, \quad \quad & \hbox{anti-periodic sector .}
\end{array}\rt.
\label{a}
\eea
The value for the periodic sector is easily understood from the fact
that we have a Bose-Fermi degeneracy in this sector. For the
anti-periodic sector we have fixed this by requiring a physical
condition, namely the open-closed duality. In this paper we have
constructed the physical Hilbert space of LCGS formalism explicitly in
terms of the pure spinor variables. The projected boundary states onto
this Hilbert space are suitable for the computation of the cylinder
diagram with manifest open-closed duality. There is no ambiguity in
the computation on the closed string side. Hence it gives a unique
answer which has to be consistent with the open string channel
computation. This fixes the $L_0$ eigenvalue of the unique ground
state in the NS sector.  

\sectiono{Massless Vertex Operators and Supersymmetry}
\label{a:massless}

Physical vertex operators in the unintegrated form are given by
certain ghost
number one operators in the BRST cohomology. The super-Poincar\'e
invariant
massless vertex operator is given by: $\lam^{\al}(z) A_{\al}(X(z),
\theta(z))$
where the function $A_{\al}(x,\theta)$ is the spinor potential for
D=10, N=1 super-Maxwell theory satisfying
$D_{\al}(\g^{\mu_1\cdots \mu_5})^{\al\bt} A_{\bt}=0$ for any
$\mu_1,\cdots \mu_5$ and $D_{\al}={\partial \over \partial
\theta^{\al}} + {1\over 2} \gb^{\mu}_{\al \bt} \theta^{\bt}
\partial_{\mu}$. The gauge invariance is given by:
$\delta A_{\al}=D_{\al} \Omega$. Using this gauge invariance
the massless vertex operators can be given the following form
in momentum space,
\bea
u(k,z) = a_{\mu}(k) b^{\mu}(k,z) + \xi^{\al}(k) f_{\al}(k,z)~,
\label{ukz}
\eea
where,
\bea
b^{\mu}(k,z) &=& {1\over 2}\lt( \lam(z) \gb^{\mu} \theta(z) \rt)
e^{ik.X}(z) + \cdots~, \cr
f_{\al}(k,z) &=& {1\over 3} \lt( \lam(z) \gb^{\mu} \theta(z)\rt)
\lt(\gb_{\mu}\theta(z)\rt)_{\al}e^{ik.X}(z) + \cdots ~,
\label{bmu-fal}
\eea
where the dots refer to terms higher order in
$\theta$. Notice that the gluon vertex operator $b^{\mu}(k,z)$ is
world-sheet fermionic as it contains terms with odd $\theta$-charges
only. Gluino vertex operator $f_{\al}(k,z)$, on the other hand, has
even $\theta$-charge,  hence is bosonic. Equation of motion and the
residual gauge invariance
are given by,
\bea
&& k^2 =0~, \quad k_{\mu}a^{\mu}(k) =0~, \quad
k_{\mu}\lt(\gb^{\mu}\xi(k) \rt)_{\al}=0~,\cr
&& \delta a^{\mu}(k) = \Lambda(k) k^{\mu}~, \quad \delta \xi^{\al}(k) =0~.
\label{on-shell}
\eea
The vertex operators in (\ref{bmu-fal}) are BRST closed when the above
equations of motion are satisfied,
\bea
\{Q_B, b^{\mu}(k,z)\} = 0~, \quad [Q_B, f_{\alpha}(k,z)] =0~.
\label{BRST-bmu-fal}
\eea
The on-shell supersymmetry transformations are (up to BRST exact terms),
\bea
\lt\{Q_{\al}, b^{\mu}(k,z) \rt\} = -{i\over 2} k_{\nu} \gb^{\mu
  \nu~\bt}_{~~\al} f_{\bt}(k,z)~, \quad
\lt[Q_{\al}, f_{\bt}(k,z) \rt] = -\gb^{\mu}_{\al \bt} b_{\mu}(k,z)~.
\label{susy-unint}
\eea
The integrated vertex operators have ghost number zero and satisfy the
following relations with the unintegrated ones,
\bea
\lt[Q_B, {\cal B}^{\mu}(k, z) \rt] =\partial b^{\mu}(k, z)~, \quad
\lt\{ Q_B, {\cal F}_{\alpha}(k, z) \rt\} = \partial f_{\alpha}(k, z)~.
\label{BRST-Bmu-Fal}
\eea
The on shell supersymmetry transformations take the similar form as in
eqs.(\ref{susy-unint}) and are given, up to total derivative terms,
by,
\bea
\lt[Q_{\al}, {\cal B}^{\mu}(k,z) \rt] = {i\over 2} k_{\nu} \gb^{\mu
  \nu~\bt}_{~~\al} {\cal F}_{\bt}(k,z)~, \quad
\lt\{Q_{\al}, {\cal F}_{\bt}(k,z) \rt\} = \gb^{\mu}_{\al \bt} {\cal
  B}_{\mu}(k,z)~.
\label{susy-int}
\eea
Clearly ${\cal B}^{\mu}(k, z)$ and ${\cal F}_{\alpha}(k, z)$ have
$\theta$ expansions with only even and odd order terms
respectively. To justify our construction of the DDF operators we need
explicit expressions for only up to first order terms.
\bea
{\cal B}^{\mu}(k, z) &=& \lt( \partial X^{\mu}(z) + i k_{\nu}
L^{\nu \mu}(z) \rt) e^{ik.X}(z) + \cdots ~, \cr
{\cal F}_{\alpha}(k, z) &=& p_{\alpha}(z)e^{ik.X}(z) \cr
&& + \lt(\gb^{\mu}\theta(z)\rt)_{\alpha}
\lt[ {1\over 2} \partial X_{\mu}(z) + ik^{\nu} \lt(N_{\nu \mu}(z)
+{1\over 2} M_{\nu \mu}(z) \rt)\rt] e^{ik.X}(z) + \cdots ~,
\label{Bmu-Fal}
\eea
where $L^{\mu \nu}=M^{\mu \nu} + N^{\mu \nu}$ is the fermionic matter
and pure spinor ghost contribution to the $SO(9,1)$ Lorentz current at
level $1$. This should be identified with the fermonic matter
contribution to the Lorentz current in NSR formalism.
Notice that at zero momentum $\th$-expansion of these operators
simplify. Using necessary OPE's it can be argued that with,
\bea
\B^{\mu}(0,z) = \del X^{\mu}(z)~, \quad \F_{\alpha}(0,z) = q_{\alpha}(z)~,
\label{Bmu-Fal-zero-mom}
\eea
one can satisfy both eqs.(\ref{BRST-Bmu-Fal}) and (\ref{BRST-bmu-fal})
with \cite{superparticle},
\bea
b^{\mu}(0,z) = {1\over 2} \lt(\lambda(z) \bar \gamma^{\mu}
\th(z)\rt)~, \quad f_{\alpha}(0,z)={1\over 3}\lt(\lambda(z) \bar
\gamma^{\mu}\th(z)\rt) \lt(\bar \gamma_{\mu}\th(z)\rt)_{\alpha}~.
\label{bmu-fal-zeromom}
\eea
This result is crucial to show eqs.(\ref{S0}).

We shall now show that
the first order term in the $\theta$-expansion
of ${\cal F}_{\alpha}(k,z)$ is as given in
eq.(\ref{Bmu-Fal}). Writing,
\bea
{\cal B}_{\mu}(k, z) &=& {\cal B}^{(0)}_{\mu}(k, z) +
{\cal B}^{(2)}_{\mu}(k, z) +\cdots ~, \cr {\cal F}_{\alpha}(k, z) &=&
{\cal F}^{(-1)}_{\alpha}(k, z) + {\cal F}^{(1)}_{\alpha}(k, z) + \cdots ~, \cr
q_{\al}(z) &=& q_{\al}^{(-1)}(z) + q_{\al}^{(1)}(z) + q_{\al}^{(3)}(z)~,
\label{theta-expand}
\eea
with the integers appearing in the superscripts of various term
referring to the $\theta$-charge and using the supersymmetry
transformations (\ref{susy-int}) one concludes (up to possible total
derivative terms),
\bea
\hbox{Res}_{z\to w} \lt[ q_{\al}^{(-1)}(z)
  \xi^{\bt}(k)\F_{\bt}^{(1)}(k,w) + q^{(1)}_{\al}(z) \xi^{\bt}(k)
\F_{\bt}^{(-1)}(k,w) \rt] &=& -\lt(\gb^{\mu}\xi(k)\rt)_{\al}
\B_{\mu}^{(0)}(k,w)~, \cr &&
\label{Res-q-F1}
\eea
where $\xi(k)$ satisfies the on-shell condition in
(\ref{on-shell}). Reading out $q^{(1)}(z)$ and $\F^{(-1)}_{\al}(k,z)$
from eqs.(\ref{susy-charge}) and (\ref{Bmu-Fal}) one shows,
\bea
\hbox{Res}_{z\to w} q^{(1)}_{\al}(z) \xi^{\bt}(k)\F^{(-1)}_{\bt}(k,w) &=&
-{1\over 2} \lt(\gb^{\mu}\xi(k)\rt)_{\al} \partial X_{\mu}(w) e^{ik.X}(w) \cr
&& + {i\over 2}k_{\mu}\xi^{\bt}(k) \lt(\gb^{\mu}\theta(w)\rt)_{\al}
p_{\bt}(w) e^{ik.X}(w)~.
\label{Res-q-F2}
\eea
Using the on-shell condition: $k_{\mu}\gb^{\mu}_{\al\bt} \xi^{\bt}(k)
=0$
and the gamma matrix property: $\eta_{\mu \nu}\gb^{\mu}_{(\al \bt}
\gb^{\nu}_{\g)\delta} =0$, one can do a manipulation to write,
\bea
k_{\mu}\xi^{\bt} \lt(\gb^{\mu}\theta \rt)_{\al} p_{\bt} = k_{\mu}
\lt(\gb_{\nu}\xi\rt)_{\al} M^{\nu \mu} + {1\over 2} k_{\mu}
\lt(\theta \gb_{\nu}\xi\rt)\lt(\gb^{\nu \mu}p\rt)_{\al}~.
\label{manipulate}
\eea
Using this and reading out the expression for $\B^{(0)}_{\mu}(k,z)$
from eqs.(\ref{Bmu-Fal}) one can write,
\bea
\hbox{Res}_{z\to w} p_{\al}(z) \xi^{\bt}(k) {\cal G}_{\bt}^{(1)}(k,w) &=&
-{1\over 2}\lt(\gb^{\mu}\xi(k)\rt)_{\al}\partial X_{\mu}(w) - i
\lt( \gb^{\mu}\xi(k)\rt)_{\al} k^{\nu} N_{\nu \mu}(w) \cr
&& -{i\over 2} \lt(\gb^{\mu}\xi(k)\rt)k^{\nu} M_{\nu \mu}(w) +
{i\over 4} \lt(\theta(w)\gb^{\mu}\xi(w)\rt) k^{\nu}
\lt(\gb_{\nu \mu}p\rt)_{\al}~. \cr &&
\label{Res-p-G}
\eea
where we have written,
\bea
\F_{\al}^{(1)}(k,z) = {\cal G}^{(1)}_{\al}(k,z) e^{ik.X}(z)~.
\label{F-G}
\eea
It can be explicitly checked that the expression for ${\cal
  G}^{(1)}_{\al}(z)$ as read from eqs.(\ref{F-G}) and (\ref{Bmu-Fal})
indeed satisfies eq.(\ref{Res-p-G}). We should also check the consistency of this result with the BRST property. The following equation,
\bea
[Q_B^{(-1)}, \xi^{\alpha}(k) \F^{(1)}_{\alpha}(k,z)] +
[Q_B^{(1)}, \xi^{\alpha}(k) \F^{(-1)}_{\alpha}(k,z)] =0~,
\label{BRST-Fal}
\eea
which is obtained by expanding the second equation in
(\ref{BRST-Bmu-Fal}) in powers of $\th$, needs to be satisfied.
Reading out the expressions for $q_B^{(-1)}(z)$ and $q_B^{(1)}(z)$ from eq.(\ref{QB}) one first derives,
\bea
\hbox{Res}_{z\to w}q_B^{(-1)}(z)\F^{(1)}_{\alpha}(k,w) &=&
\lt(\bar \gamma^{\mu} \lambda(w)\rt)_{\alpha} \lt[{1\over 2} \partial X_{\mu}(w) + i k^{\nu} \lt(N_{\nu \mu}(w) + {1\over 2}M_{\nu \mu}(w) \rt) \rt] e^{ik.X}(w) \cr
&& -{i\over 4} k^{\nu} \lt(\bar \gamma^{\mu} \th(w) \rt)_{\alpha} \lt(\lambda(w) \bar \gamma_{\nu \mu}p(w) \rt) e^{ik.X}(w)~, \cr && \cr
\hbox{Res}_{z\to w}q_B^{(1)}(z)\F^{(-1)}_{\alpha}(k,w) &=& \lt[ -{1\over 2} \lt( \bar \gamma^{\mu} \lambda(w)\rt)_{\alpha} \partial X_{\mu}(w) + \rt. \cr
&& \lt. {i\over 2} k_{\mu} \lt(\bar \gamma^{\mu}\partial \lambda (w) \rt)_{\alpha} + {i\over 2} k_{\mu} \lt( \lambda(w) \bar \gamma^{\mu} \th(w) \rt) p_{\alpha}(w)\rt] e^{ik.X}(w)~. \cr &&
\eea
Then using the on-shell condition for $\xi^{\alpha}(k)$ and the result (\ref{manipulate}) one shows that the condition (\ref{BRST-Fal}) is indeed satisfied. One may wonder what happens to the $N_{\mu \nu}$ dependent term in the first equation as there is no other term that can cancel it. This term can be shown to drop off by using the following identity,
\bea
N^{\mu \nu}\lt( \bar \gamma_{\nu} \lambda \rt)_{\alpha} = -{1\over 4} \lt( w\gamma^{\mu} \bar \gamma^{\nu}\rt)_{\alpha} \lt(\lambda \bar \gamma_{\nu}\lambda \rt) - {1\over 2} \lt( \bar \gamma^{\mu}\lambda \rt)_{\alpha} \lt( w\lambda \rt)~,
\eea
and the on-shell condition for $\xi^{\alpha}(k)$.

\sectiono{The {\it Absence of Maximal Left-Moving $\th$-Charges} Theorem}
\label{a:theorem}

Below we state and prove the {\it absence of maximal left-moving $\th$-charges} theorem.
\bea
&\hbox{\bf Statement}& \cr && \cr
&\begin{array}{l}
\hbox{\it The $n$-th order terms $\B^{(n)I}(k^-,z)$ and $\F^{(n)a}_L(k^-,z)$
in the DDF vertex operators}\cr
\hbox{\it $\B^I(k^-,z)$ and $\F^a_L(k^-,z)$ respectively
do not contain terms with charge $(n,0)$ for}\cr
\hbox{\it $n>1$ and $(n+1,-1)$ for $n>-1$~.}
\end{array} & \cr &&
\label{theorem} 
\eea 
\begin{center}
{\bf Proof}
\end{center}
We start by proving that ${\cal O}^{(n)}(z)$ does not have a term with charge
$(n+1,-1)$ for $n>-1$, where ${\cal O}^{(n)}(z)$ stands for either $\B^{(n)I}(k^-,z)$ or
$\F^{(n)a}_L(k^-,z)$. To do that let us consider the two commutation relations involving 
$Q^a_L$ in eqs.(\ref{susy-BI}) and (\ref{susy-Fal}) for non-zero $k^-$. These relations
imply,
\bea
\hbox{Res}_{z\to w} \lt[q_L^{(-1,0)}(z) {\cal O}^{(n)}(w) +
  \lt\{q_L^{(0,1)}(z)+ q_L^{(1,0)}(z)\rt\} {\cal O}^{(n-2)}(w) +
  q_L^{(1,2)}(z) {\cal O}^{(n-4)}(w) \rt] =0~, \cr
\label{susyL-On}
\eea
where,
\bea
&& q_L^{(-1,0)} = p_L~, \quad q_L^{(0,1)} = {1\over 2} \del X^I
\sigma^I \th_R~, \quad q_L^{(1,0)} = {1\over \sqrt{2}} \del X^+
\th_L~, \cr
&& q_L^{(1,2)}= -{1\over 12} (\th_R\del \th_R) \th_L
+{1\over 24} \lt\{(\th_L\sigma^I\del \th_R) + (\th_R \bar \sigma^I\del
\th_L)\rt\} \sigma^I\th_R~.
\label{qL(p,q)}
\eea
If ${\cal  O}^{(n)}$ contains a term with charge $(n+1,-1)$ then the first term
on the left hand side of eq.(\ref{susyL-On}) will produce a term with charge $(n,-1)$.
\bea
\hbox{Res}_{z\to w} q_L^{(-1,0)}(z) {\cal O}^{(n)}(w) \rightarrow
{\cal L}^{(n,-1)}(w)~,
\label{Ln,-11}
\eea
which needs to be cancelled by similar contributions coming from the rest of the terms. 
There can not be any contribution coming from the last term. This is because the only
operator of negative left (right)-charge is $p_L$ ($p_R$) which has left-charge 
(right-charge) $-1$ and dimension $1$ and ${\cal O}^{(n)}$ has dimension $1$ for any $n$.  
Therefore,
\bea
\hbox{Res}_{z\to w} \lt[q^{(0,1)}_L(z) {\cal O}^{(n-2)}(w) +
  q^{(1,0)}(z) {\cal O}^{(n-2)}(w)  \rt] \rightarrow - {\cal
  L}^{(n,-1)}(w)~.
\label{Ln,-12}
\eea
In order for the first term to contribute to the right hand side
${\cal O}^{(n-2)}$ needs to have a term with charge $(n,-2)$, which is
not possible for the same reason described above. Also the second term can not 
contribute, because $q^{(1,0)}_L$ does not have a residue with a dimension one term with
charge $(n-1,-1)$. Therefore we must have,
\bea
{\cal L}^{(n,-1)}(z) = 0~,
\eea
which implies ${\cal O}^{(n)}$ can not have a term with charge $(n+1,-1)$ for $n>-1$. 
The above argument is invalid for $n=-1$. This is because the right side of 
eq.(\ref{Ln,-11}) is trivial and the last two terms on the left side of 
eq.(\ref{susyL-On}) do not exist. Therefore ${\cal O}^{(-1)}$ can have a term with 
charge $(0,-1)$ while satisfying eq.(\ref{susyL-On}). 

Let us now turn to prove the other part of the theorem, namely the term with charge 
$(n,0)$ does not appear in ${\cal O}^{(n)}$ for $n>1$. If ${\cal O}^{(n)}$ has a term 
with charge $(n,0)$ then the first term in eq.(\ref{susyL-On}) will produce a term 
with charge $(n-1,0)$
\bea
\hbox{Res}_{z\to w} q_L^{(-1,0)}(z) {\cal O}^{(n)}(w) \rightarrow
     {\cal K}^{(n-1,0)}(w)~,
\label{Kn-1,01}
\eea
which needs to be cancelled by a similar term produced by the last two
terms in eq.(\ref{susyL-On}). As argued previously the last term in 
eq.(\ref{susyL-On}) can not produce such a term. This implies,
\bea
\hbox{Res}_{z\to w} \lt[ q_L^{(0,1)}(z) {\cal O}^{(n-2)}(w) +
  q_L^{(1,0)}(z){\cal O}^{(n-2)}(w) \rt] \rightarrow - {\cal
  K}^{(n-1,0)}(w)~.
\label{Kn-1,02}
\eea
Let us first consider the first term on the left hand side of
(\ref{Kn-1,02}). In order for this term to contribute to the right
hand side ${\cal O}^{(n-2)}$ should necessarily have a term with
charge $(n-1, -1)$. Recalling that we are considering $n>1$, this requirement
can not be satisfied due to the part of the theorem that has been proved first.
We now consider the second term on the left hand side. From
eqs.(\ref{qL(p,q)}) it is easy to argue that in order for the second
term to produce an $(n-1,0)$ term, it is necessary that ${\cal
  O}^{(n-2)}$ have an $(n-2, 0)$ term with a factor of $\del
X^-$. Below we shall argue that the following statement is true:
\bea
&& \hbox{{\it The DDF vertices $\B^I(k^-,z)$ and $\F_L^a(k^-,z)$ do
    not contain the operator $\del X^-(z)$}} \cr
&& \hbox{{\it in their $\th$-expansion.}}
\label{result1}
\eea
Assuming this result for the time being we conclude that the second term 
on the left hand side of (\ref{Kn-1,02}) does not contribute to the 
right hand side. This implies,
\bea
{\cal K}^{(n-1,0)}(z) =0~,
\label{Kn-1,03}
\eea
which implies ${\cal O}^{(n)}(z)$ does not have a term with charge
$(n,0)$ for $n>1$. For $n=0,-1$ the first term in eq.(\ref{susyL-On}) 
gives zero for $(0,0)$ and $(-1,0)$ terms coming from ${\cal O}^{(0)}$ and
${\cal O}^{(-1)}$ respectively and the rest of terms are nonexistent in both 
the cases. Therefore ${\cal O}^{(0)}$ and ${\cal O}^{(-1)}$ can have terms with
charges $(0,0)$ and $(-1,0)$ respectively while satisfying eq.(\ref{susyL-On}). 

We now proceed to prove the result (\ref{result1}). 
Let us first consider $\B^I(k^-,z)$. Any term which will give rise to
$\del X^-(z)$ in $\B^I(k^-,z)$ should come from a covariant term of
the following form in $\B^{\mu}(k,z)$: $\del X_{\nu}(z) {\cal A}^{\mu
  \nu}(k,z)$, where ${\cal A}^{\mu \nu}(k,z)$ is a dimension zero
operator in the fermionic matter sector, and therefore constructed
entirely out of $\th$'s. The simplest possibility is $k^{\mu}k^{\nu}$
which does not have any $\th$. For the momentum restriction relevant
for $\B^I(k^-,z)$, this gives rise to $\del X^+$, not $\del X^-$. To 
look for terms with non-zero number of $\th$'s we should keep in
mind that we must have even number of $\th$'s in a given term and
that, because the gamma matrices are symmetric, only a third rank
tensor $(\th \bar \gamma^{\mu \nu \rho} \th)$ (which will be called
$\th^2$ hereafter) can be constructed out of two $\th$'s. Therefore an
eligible term will be a product of such third rank tensors and
momenta. The two vector indices in ${\cal A}^{\mu \nu}(k,z)$ can, in
general, come from any such factors. It is easy to see that if any of
them comes from momentum then the term either does not contribute to
$\B^I(k^-,z)$ at all  or gives rise to $\del X^+$, not $\del
X^-$. Also the momentum independent terms can be ignored as we know from
eq.(\ref{Bmu-Fal-zero-mom}) that at zero momentum there is no $\del
X^-$ in $\B^I(k^-,z)$. The other possibilities include two cases where
both the vector indices come from the same $\th^2$ factor and two different
$\th^2$ factors. In the first case we have one vector index from the
relevant $\th^2$ factor which is contracted with another $\th^2$ factor
or monemtum. In the second case each of the two relevant $\th^2$
factors will have two vector indices contracted with other $\th^2$
factors and/or momenta. In order to have a $\del X^-$ in $\B^I(k^-,z)$
one of the vector index has to be $+$. Therefore we have a situation
where we need to have a $\th^2$ factor with one vector index to be $+$
and one or two (depending on the cases described above) other vector
indices to be contracted with other $\th^2$ factors and/or momenta. It
is easy get convinced that a full contraction of these kinds will
always involve momentum contraction(s). Since the only nonzero
component of momentum is $k^-$ these momentum contractions will always
induce a $+$ index in the original $\th^2$ factor which had a free $+$
index. Since the $\th^2$ factor is antisymmetric in its indices this
must be zero. This establishes that $\del X^-(z)$ does not appear in 
$\B^I(k^-,z)$.

Let us now consider $\F_L^a(k^-,z)$. The covariant term in
$\F_{\alpha}(k,z)$ that will potentially give rise to $\del X^-$ in
$\F_L^a(k^-,z)$ should have the form: $\del X_{\mu} {\cal
  D}^{\mu}_{\alpha}(k,z)$, where ${\cal D}^{\mu}_{\alpha}(k,z)$ is a
dimensionless operator constructed entirely out of $\th$'s. The
possibilities are as follows:
\bea
\begin{array}{cc}
\hbox{Class I} & \hbox{Class II} \cr & \cr
\lt(\bar \gamma^{\mu}\th \rt)_{\alpha} 
{\cal E}_0(k,z) & 
\lt(\bar \gamma_{\rho}\th \rt)_{\alpha} {\cal E}_1^{\mu \rho}(k,z) \cr
\lt(\bar \gamma^{\mu}_{~\rho_1 \rho_2}\th \rt)_{\alpha} 
{\cal E}_2^{[\rho_1 \rho_2]}(k,z) &
 \lt(\bar \gamma_{\rho_1 \rho_2 \rho_3}\th \rt)_{\alpha} 
{\cal E}_3^{\mu [\rho_1 \rho_2 \rho_3]}(k,z) \cr
\lt(\bar \gamma^{\mu}_{~\rho_1 \rho_2 \rho_3 \rho_4}\th \rt)_{\alpha}
{\cal E}_4^{[\rho_1 \rho_2 \rho_3 \rho_4]}(k,z) \quad & \quad 
\lt( \bar \gamma_{\rho_1 \rho_2 \rho_3 \rho_4 \rho_5} 
\th \rt)_{\alpha} {\cal E}_5^{\mu [\rho_1 \rho_2 \rho_3 
\rho_4 \rho_5]}(k,z)
\end{array}
\eea
where all the operators denoted by ${\cal E}$ with various tensor
structures are products of $\th^2$ terms and momenta. None of the
class I operators appears in $\F_L^a(k^-,z)$ when the free vector
index is set to $+$. This is simply because the prefactor linear in
$\th$ that appears in each of these operators is projected to the
``wrong'' chirality. Although similar projection gives the ``right''
chirality for the class II operators, they do not appear because of
the momentum restriction involved in the ${\cal E}$ operators. Each of 
the ${\cal E}$ operators involves a $\th^2$ factor whose one vector index   
is set to $+$ and one or two other indices are contracted to other
$\th^2$ factors and/or momenta. We have argued before that such terms
are zero. This establishes that $\del X^-(z)$ does not appear
in $\F_L^a(k^-,z)$.

\sectiono{The Instantonic D-Branes}
\label{a:Dinstanton}

Here we shall discuss the boundary conditions and boundary states for both
the BPS and non-BPS instantonic D-branes in type IIB string theory. Following
\cite{mukhopadhyay05} we shall work in the free CFT.

In the BPS case, boundary condition for the bosonic matter
part of the CFT is , as usual, given by (on UHP),
\bea
\partial X^{\mu}(z) = - (M^V)^{\mu}_{~\nu} \bar \partial X^{\nu}(\bar
z) ~, \quad \hbox{ at } z=\bar z~,
\label{Xbc}
\eea
where $M^V$ is the diagonal reflection matrix with $-1$ for the
Neumann directions and $+1$ for the Dirichlet directions. For the
fermionic matter and bosonic ghost sectors the boundary conditions
can be obtained by demanding that the scalars and vectors constructed
out of the fields in these sectors are related at the boundary in the
following way,
\bea
\Phi(z) = \tilde \Phi(\zb)~, \quad A^{\mu}(z) = -(M^V)^{\mu}_{~\nu}
\tilde A^{\nu}(\zb)~.
\label{scalar-vector-bc}
\eea
The result is,
\bea
U^{\al}(z) = -i \eta (M^S)^{\al}_{~\bt}\tilde U^{\bt}(\zb)~,
\quad V_{\al}(z) = i \eta (\bar M^S)_{\al}^{~\bt}\tilde V_{\bt}(\zb)
~, \quad \hbox{ at } z=\zb~,
\label{UVbc-BPS}
\eea
where,
\bea
M^S = \g^{I_1I_2\cdots I_{p+1}}~, \quad \bar M^S =\gb^{I_1I_2\cdots I_{p+1}}~,
\label{MSMbS}
\eea
with $p$ being odd and $I_1, I_2, \cdots I_{p+1}$ being the Neumann
directions (all spatial). $\eta=\pm 1$ correspond to brane and
anti-brane.
Since the matrices $M^S$ and $\bar M^S$ includes only the spatial
directions we have the following properties,
\bea
M^S (\bar M^S)^T &=& \Iop_{16}~, \cr
M^S \g^{\mu} (M^S)^T &=&(M^V)^{\mu}_{~\nu} \g^{\nu}~, \quad (M^S)^T
\gb^{\mu} M^S =(M^V)^{\mu}_{~\nu} \gb^{\nu}~, \cr
\bar M^S \gb^{\mu} (\bar M^S)^T &=&(M^V)^{\mu}_{~\nu} \gb^{\nu}~, \quad
(\bar M^S)^T \g^{\mu} \bar M^S =(M^V)^{\mu}_{~\nu} \g^{\nu}~.
\label{MSMbS-prop}
\eea
All the above equations differ by a sign with respect to the case
where the matrices include the time direction as in
\cite{mukhopadhyay05}. The BRST and supersymmetry currents are related
on the boundary in the following way,
\bea
j_B(z) = \tilde j_B(\zb)~, \quad q_{\al}(z) = i\eta (\bar
M^S)_{\al}^{~\bt} \tilde q_{\bt}(\zb)~.
\label{BRST-susy-bc}
\eea
As a result the BPS boundary state $|\hbox{Inst}_p, \eta \ra_{BPS}$ is
BRST invariant and preserves the expected combination of the supersymmetry,
\bea
\lt(Q_B + \tilde Q_B\rt) |\hbox{Inst}_p,\eta \ra_{BPS} =0~, \quad
\lt(Q_{\al}+i\eta (\bar M^S)_{\al}^{~\bt}\tilde Q_{\bt}\rt)
|\hbox{Inst}_p,\eta\ra_{BPS}=0~.
\label{BRST-susy-gluing}
\eea

As we have seen in \cite{mukhopadhyay04}, unlike the case of BPS
D-branes, open string boundary conditions for non-BPS D-branes do not involve
the spinor matrices representing reflections along Neumann
directions. Therefore these boundary conditions, once written in terms
of the vector matrix $M^V$, should look the same for both Lorentzian
and instantonic D-branes. Indeed the bosonic matter and combined
fermionic matter and bosonic ghost parts of the non-BPS D-instanton
boundary conditions are given by eqs.(\ref{Xbc}, \ref{bcnBPS})
respectively with $M^V$ representing reflections along the Neumann
directions of the considered non-BPS D-brane.

The boundary states for both the BPS and non-BPS instantonic D-branes
can be constructed explicitly in terms of the oscillators, as was done
in \cite{mukhopadhyay05}, but we do not need the explicit expression
for the purpose of the present paper. All we need is to argue using
the boundary conditions (\ref{UVbc-BPS}) and equations
(\ref{MSMbS-prop}) that any holomorphic spinor with either upper or
lower spinor index will be related to the corresponding anti-holomorphic one
following the same rule as followed in eqs.(\ref{UVbc-BPS}).

\end{document}